\documentclass[prb,showpacs,preprintnumbers,amsmath,amssymb]{revtex4} %draft,aps,
\usepackage{graphicx}% Include figure files

\begin{document}

%%\draft
\title{Critical fluctuations and pseudogap observed in the microwave
conductivity of BSCCO and YBCO thin films}
\author{
D.-N.~Peligrad$^{*}$, and M.~Mehring}
\address{2. Physikalisches Institut, Universit\"at Stuttgart,
70550 Stuttgart, Germany}
\author{
A.~Dul\v{c}i\'{c}}
\address{Department of Physics, Faculty of Science,
University of Zagreb, POB 331, 10002 Zagreb, Croatia}
\date{submitted: July 20, 2003; received on {\it condmat} \today}

\thanks{present address: Philips Research
Laboratories, Weisshausstrasse 2, D-52066, Aachen, Germany}

\begin{abstract}
Critical fluctuations have been studied in the microwave
conductivity of $Bi_2Sr_2CaCu_2O_{8+\delta}$,
$Bi_2Sr_2Ca_2Cu_3O_{10+\delta}$, and $YBa_2Cu_3O_{7-\delta}$ thin
films above $T_c$. It is found that a consistent analysis of the
real and imaginary parts of the fluctuation conductivity can be
achieved only if an appropriate wavevector or energy cutoff in the
fluctuation spectrum is taken into account. In all of the three
underdoped superconducting films one observes strong fluctuations
extending far above $T_c$. The coherence length inferred from the
imaginary part of the conductivity exhibits the static critical
exponent $\nu = 1$ very close to $T_c$, and a crossover to the
region with $\nu = 2/3$ at higher temperatures. In parallel, our
analysis reveals the absence of the normal conductivity near
$T_c$, i.~e. fully opened pseudogap. Following the crossover to
the region with $\nu = 2/3$, the normal conductivity is gradually
recovered, i.~e. the closing of the pseudogap is monitored.\\
\end{abstract}

 \maketitle

\section{INTRODUCTION} \label{Sect1}
Soon after the discovery of high-$T_c$ superconductors it was
found that fluctuations were more pronounced in these compounds
than in the classical low temperature superconductors
\cite{Freitas:87,Ausloos:88}. It is due primarily to short
coherence lengths and large transition temperatures, but
anisotropy and the degree of doping may also play a role. While
classical low temperature superconductors exhibited only Gaussian
type fluctuations, \cite{Skocpol:75} it was estimated that
high-$T_c$ superconductors could give experimental access to a
study of critical fluctuations \cite{Lobb:87}. Yet, early studies
of the $dc$ fluctuation conductivity
\cite{Hopfengartner:91,Gauzzi:95} were interpreted within the
Gaussian behavior, except for a narrow region (fraction of a
degree) around $T_c$ which was left out as possibly critical.
Later on, the penetration depth measurements in
$YBa_2Cu_3O_{7-\delta}$ revealed critical behavior as wide as $\pm
10$K from $T_c$, \cite{Kamal:94} belonging to the universality
class 3D XY. More recent measurements of thermal expansivity
\cite{Pasler:98,Meingast:01} and two-coil inductive measurements
\cite{Osborn:02} have confirmed this wide range of critical
fluctuations, while $dc$ fluctuatuation conductivity measurements
still claimed very narrow critical regions
\cite{Cimberle:97,Menegotto:97,Han:98,Han:00,Silva:01,Menegotto:01}.
A major problem in the analysis of a $dc$ fluctuation conductivity
is to account for the short wavelength cutoff,
\cite{Hopfengartner:91,Gauzzi:95,Cimberle:97,Silva:01} and energy
cutoff \cite{Carballeira:01,Vina:02} at higher temperatures. These
effects are so strong that other inherent properties of the
fluctuation conductivity can be obscured.
\par
A more stringent test of a given theoretical model can be made
through a study of the microwave fluctuation conductivity since it
yields two experimental curves, one for the real part
$\sigma_1(T)$, and the other for the imaginary part $\sigma_2(T)$,
with different shapes but ensuing from the same physics assumed in
the model. Some of the earlier microwave studies reported Gaussian
and critical behavior,
\cite{Anlage:96,Booth:96,Nakielski:97,Waldram:99} but failed to
account for the cutoff effects. However, a recent analysis has
shown that the short wavelength cutoff yields a strong effect at
higher temperatures, and a small, but experimentally detectable
feature at $T_c$ \cite{Peligrad:03}. The latter was shown to be
useful in setting a constraint on the cutoff parameters, which are
then used in the whole temperature region above $T_c$. Further
advantage of the $ac$ fluctuation conductivity is that the
imaginary part $\sigma_2(T)$ has no contribution from the normal
electrons so that its analysis is free from subtraction problems
often encountered in $dc$ conductivity studies.
\par
In this paper we present an analysis of the microwave fluctuation
conductivity in $Bi_2Sr_2CaCu_2O_{8+\delta}$ (BSCCO-2212),
$Bi_2Sr_2Ca_2Cu_3O_{10+\delta}$  (BSCCO-2223), and
$YBa_2Cu_3O_{7-\delta}$  (YBCO) thin films. The central result is
the temperature dependence of the coherence length which shows
multiple critical regimes in each of the superconductors. Also
shown in the present analysis is the reduction of the contribution
of the normal conductivity in the measured real part $\sigma_1(T)$
of the complex conductivity. It is a direct evidence of the
reduction of the one-electron density of states at the Fermi
level, known as the pseudogap effect \cite{Timusk:99}.
\section{THEORETICAL BACKGROUND} \label{Sect2}
In this paper, we use the expressions for the $ac$ fluctuation
conductivity which have been derived and extensively discussed
recently \cite{Peligrad:03}. The derivation was based on the time
dependent Ginzburg-Landau theory and the approach introduced by
Schmidt \cite{Schmidt:68} long time ago. The new feature was the
cutoff in the integration over the $\bf k$-space which was imposed
in order to comply with the slow variation principle of the
Ginzburg-Landau theory \cite{Tinkham:95}. Equivalent expressions
were also obtained \cite{Silva:02} from the linear response to an
applied $ac$ field \cite{Dorsey:91}.
\par
In order to treat also the critical state, we have adopted the
requirement that close to the superconducting transition all
physical quantities should be expressed through the coherence
length $\xi(T)$. In particular, the relaxation time of the lowest
fluctuation mode becomes $\tau_0 \propto \xi^z$, where z is the
dynamical critical exponent \cite{Fisher:91}. This is essential
since the temperature dependence of the coherence length in the
critical state takes a form different from that of the Gaussian or
mean field result. On the other hand, the very form of the
expressions obtained using the time dependent Ginzburg-Landau
theory remain valid even in the critical state \cite{Wickham:00}.
\par
For the three dimensional ($\it 3D$) anisotropic case the $\it ac$
fluctuation conductivity is given by \cite{Peligrad:03} %%
\begin{equation}
\displaystyle {\widetilde {\sigma}}^{\hspace{0.05cm}{\it 3D}} =
\frac{e^2}{32\hbar\xi_{0c}}\left(\frac{\xi(T)}{\xi_0}\right)^{z-1}
\int_0^{Q_{ab}}\int_{-Q_c}^{Q_c}
\frac{4\, q_{ab}^{\,3}\, \left[1 - i \Omega (1+q_{ab}^2+q_c^2)^{-1}\right]}
{\pi (1+q_{ab}^2+q_c^2)[\Omega^2+(1+q_{ab}^2+q_c^2)^2]}
\,d q_{ab}\,d q_c \,\,\, ,
\label{1}
\end{equation}
where the prefactor is the well known Aslamazov-Larkin term,
\cite{Tinkham:95} except that the Gaussian temperature dependence
$\displaystyle 1/\sqrt{\epsilon}$ is replaced by that of the
reduced coherence length $\xi(T)/\xi_0$. This dimensionless
temperature dependent function is equal for both, the in-plane and
c-axis coherence lengths, i.~e. $(\xi_{ab}(T)/\xi_{0ab}) =
(\xi_c(T)/\xi_{0c})$, \cite{Peligrad:03} so that we omit the
subscripts. The dimensionless parameter %%
\begin{equation}
\displaystyle \Omega(\omega,T)= \frac{\omega \tau_0}{2} =
\frac{\pi}{16}\frac{\hbar \omega}{k_B T}
\left(\frac{\xi(T)}{\xi_0}\right)^z \, ,
\label{2}
\end{equation}
involves the operating microwave frequency $\omega$ and the
temperature dependence of the reduced coherence length. Note that
we have used here the thermodynamically correct form $k_B T$,
rather than the approximative $k_B T_c$ used before
\cite{Peligrad:03}. The latter will be used below only in
theoretical simulations where $\xi(T)/\xi_0$ is used as the only
variable. The cutoff in the fluctuation wavevector is introduced
in $\displaystyle Q_{ab}(T) = k_{ab}^{max} \xi_{ab}(T) = \sqrt{2}
\Lambda_{ab}\,\xi_{ab}(T)/ \xi_{0ab}$ for the {\it ab}-plane and
$\displaystyle Q_c(T) = k_c^{max} \xi_c(T) = \Lambda_c\,\xi_c(T)/
\xi_{0c}$ along the {\it c}-axis, rather than a single cutoff on
the modulus \cite{Silva:02}. The analytical result of the
integration in Eq.~(\ref{1}) and its application in the data
analysis has been reported in detail \cite{Peligrad:03}. A careful
analysis has shown that, due to this cutoff, the real and
imaginary parts of the $ac$ fluctuation conductivity are not equal
at $T_c$. The ratio $\sigma_2(T_c)/\sigma_1(T_c)$, which can be
determined straightforwardly from the experimental data, puts a
constraint on the choice of the parameters $\Lambda_{ab}$ and
$\Lambda_c$ \cite{Peligrad:03}.
\par
Time dependent Ginzburg-Landau approach is based on relaxational
dynamics so that the dynamic critical exponent should be $z = 2$
\cite{Hohenberg:77}. Indeed, it has been shown by expansion of the
$\cal S$-functions in the limit of $T_c$ that only for $z = 2$ one
obtains a finite nonzero $ac$ fluctuation conductivity, in
agreement with experimental observations \cite{Peligrad:03}. This
value of the dynamic critical exponent was also found in Monte
Carlo simulations for a vortex loop model of the superconducting
transition \cite{Aji:01}.
\par
With the slow variation principle alone, the parameters
$\Lambda_{ab}$ and $\Lambda_c$ may be kept as constants from $T_c$
up to any higher temperature. However, it has been shown recently
that energy cutoff should also be imposed in order to satisfy the
uncertainty principle for the minimum wave packet of the Cooper
pairs \cite{Carballeira:01,Vina:02}. For the anisotropic $\it 3D$
case one should have the condition
\begin{equation}
\displaystyle (\xi_0/\xi(T))^2 + 2\Lambda_{ab}^2 + \Lambda_c^2 = C \, ,
\label{3}
\end{equation}
where $C$ is the energy cutoff parameter. At $T_c$ the first term
on the left hand side of Eq.~(\ref{3}) vanishes and one retrieves
the pure wavevector cutoff condition. With the constraint ensuing
from the experimental ratio $\sigma_2(T_c)/\sigma_1(T_c)$,
\cite{Peligrad:03} and Eq.~(\ref{3}) with a given energy cutoff
parameter $C$, one can determine the $\Lambda$'s at $T_c$. Within
the concept of the energy cutoff, one should keep $C$ fixed at all
temperatures. Eq.~(\ref{3}) can then be satisfied only if the
parameters $\Lambda_{ab}$ and $\Lambda_c$ become temperature
dependent. It is reasonable to assume that their ratio $K =
\Lambda_{ab}/\Lambda_c$ remains constant, i.~e. as determined at
$T_c$. Then, Eq.~(\ref{3}) yields both $\Lambda$'s as functions of
the reduced coherence length $\xi(T)/\xi_0$. With these
replacements in our expressions for the $ac$ fluctuation
conductivity \cite{Peligrad:03} one accounts also for the energy
cutoff. One may remark that the value $ C= 1$ implies that the
minimum $\xi(T)$ is equal to $\xi_0$. However, for $C < 1$ the
minimum $\xi(T)$ is bigger than $\xi_0$. This is important in
high-$T_c$ superconductors where $\xi_0$ is found to be
unphysically small \cite{Peligrad:03} so that the minimum $\xi(T)$
must be bigger than that value. It is given by min($\xi(T)$) =
$\xi_0 / \sqrt{C}$.
\par
\begin{figure}
\centerline{\includegraphics[width=0.7\textwidth]{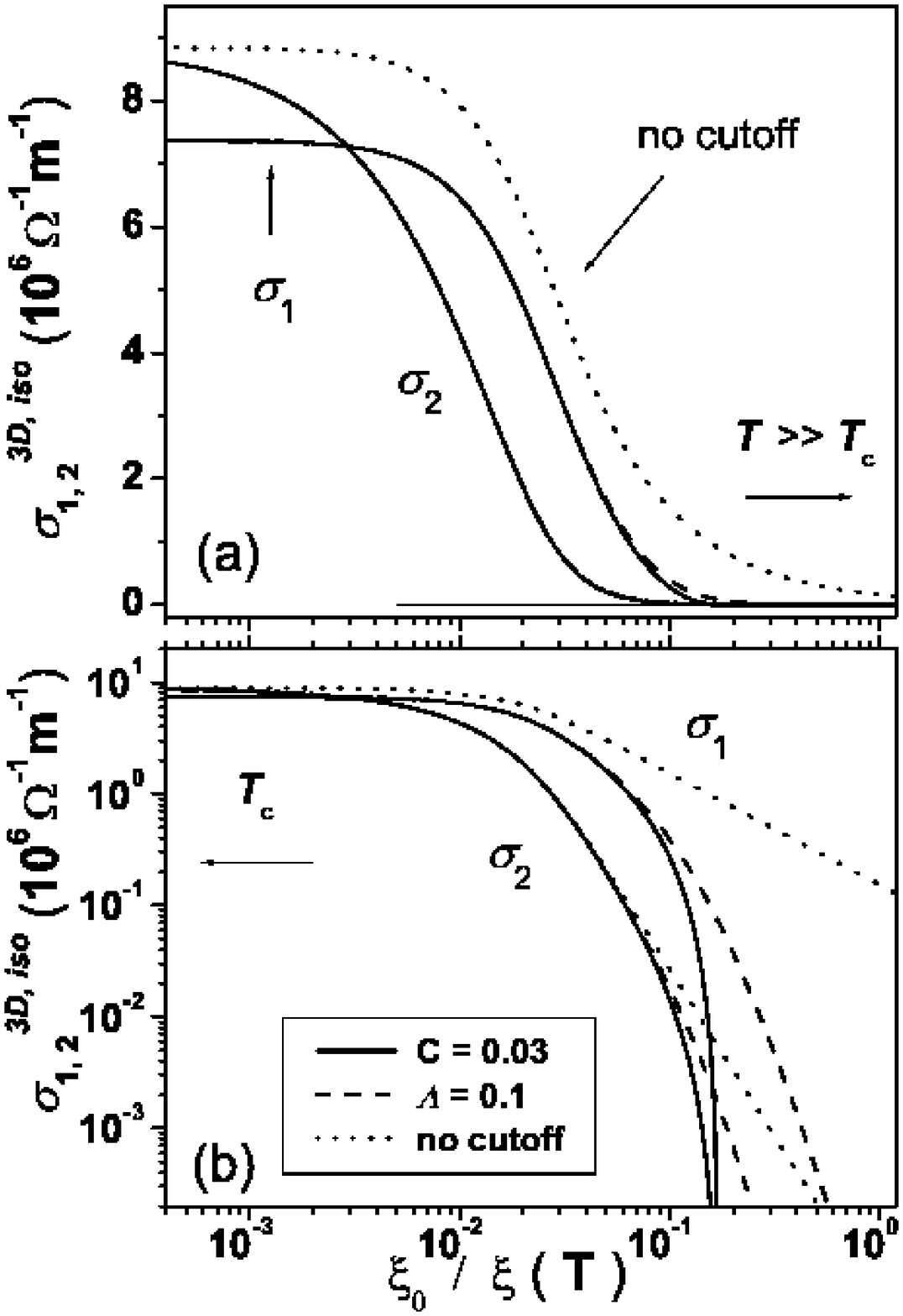}}
\caption{Calculated $\it ac$ fluctuation conductivity in the
isotropic $\it 3D$ case. The dotted lines show the results in the
case when no cutoff is taken into account. The dashed lines are
obtained with the wavevector cutoff ($\Lambda = 0.1$) consistent
with the ratio $(\sigma_2(T_c)/\sigma_1(T_c)) = 1.2$, which is a
typical experimental value. The full lines involve the energy
cutoff ($C = 0.03$) that matches the wavevector cutoff at
temperatures close to $T_c$. The upper and lower curves represent
$\sigma_1$ and $\sigma_2$, respectively.} \label{Fig1}
\end{figure}

In order to illustrate the importance of the cutoff effects in a
data analysis, we may simulate some cases. In Fig.~\ref{Fig1} we
present the calculated $\sigma$'s for an isotropic superconductor.
In that case, there is only one cutoff parameter $\Lambda$
\cite{Peligrad:03}. We have assumed that
$(\sigma_2(T_c)/\sigma_1(T_c)) = 1.2$, which is a typical
experimental value. It yields unequivocally the values $\Lambda =
0.1$, and $C = 0.03$. The $\sigma$'s in Fig.~\ref{Fig1} are
calculated as functions of $(\xi_0/\xi(T))$ in order to make the
presentation independent of the specific temperature dependence of
the coherence length. The dotted, dashed and full lines in
Fig.~\ref{Fig1} present the calculated $\sigma$'s using no cutoff,
wavevector cutoff with the value of $\Lambda$ fixed, and energy
cutoff with $C$ fixed, respectively. The upper three and lower
three curves refer to $\sigma_1$ and $\sigma_2$, respectively. One
can see that the cutoff brings about significant changes in the
calculated curves so that great care is needed in attempts to fit
a given set of experimental data.
\par
\begin{figure}
\centerline{\includegraphics[width=0.8\textwidth]{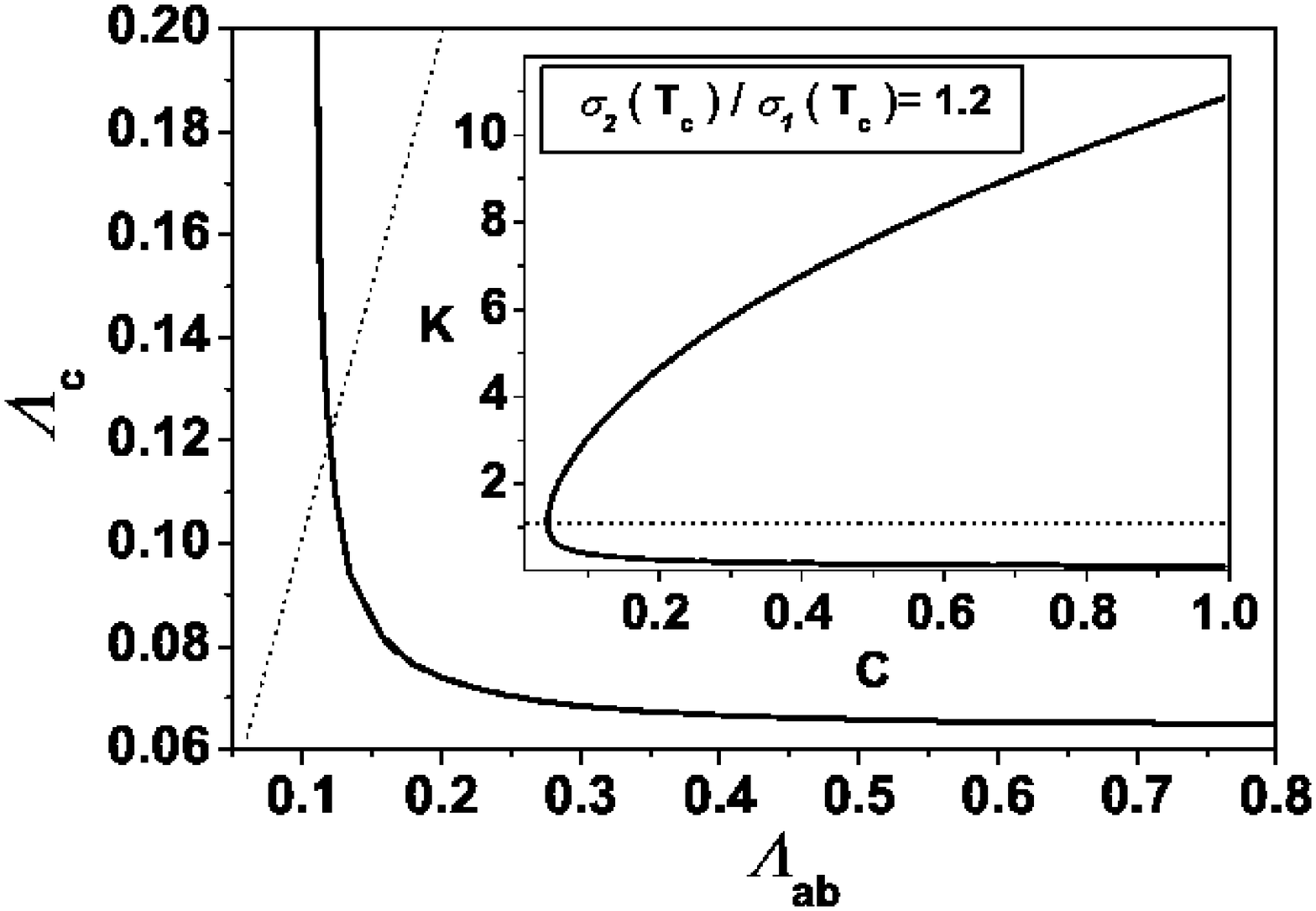}}
\caption{Possible choices of the parameters $\Lambda_{ab}$ and
$\Lambda_c$ when $(\sigma_2(T_c)/\sigma_1(T_c)) = 1.2$. The inset
shows an alternative presentation in terms of the ratio $K =
\Lambda_{ab}/\Lambda_c$ and the energy cutoff parameter $C$. The
lower branch of this curve corresponds to the cases $\Lambda_{ab}
\ll \Lambda_c$ while the upper branch is for $\Lambda_{ab} \gg
\Lambda_c$. The crossover $\Lambda_{ab} = \Lambda_c$ is shown by
the dotted line in the main panel, and $K = 1$ in the inset.}
\label{Fig2}
\end{figure}
\begin{figure}
\centerline{\includegraphics[width=0.65\textwidth]{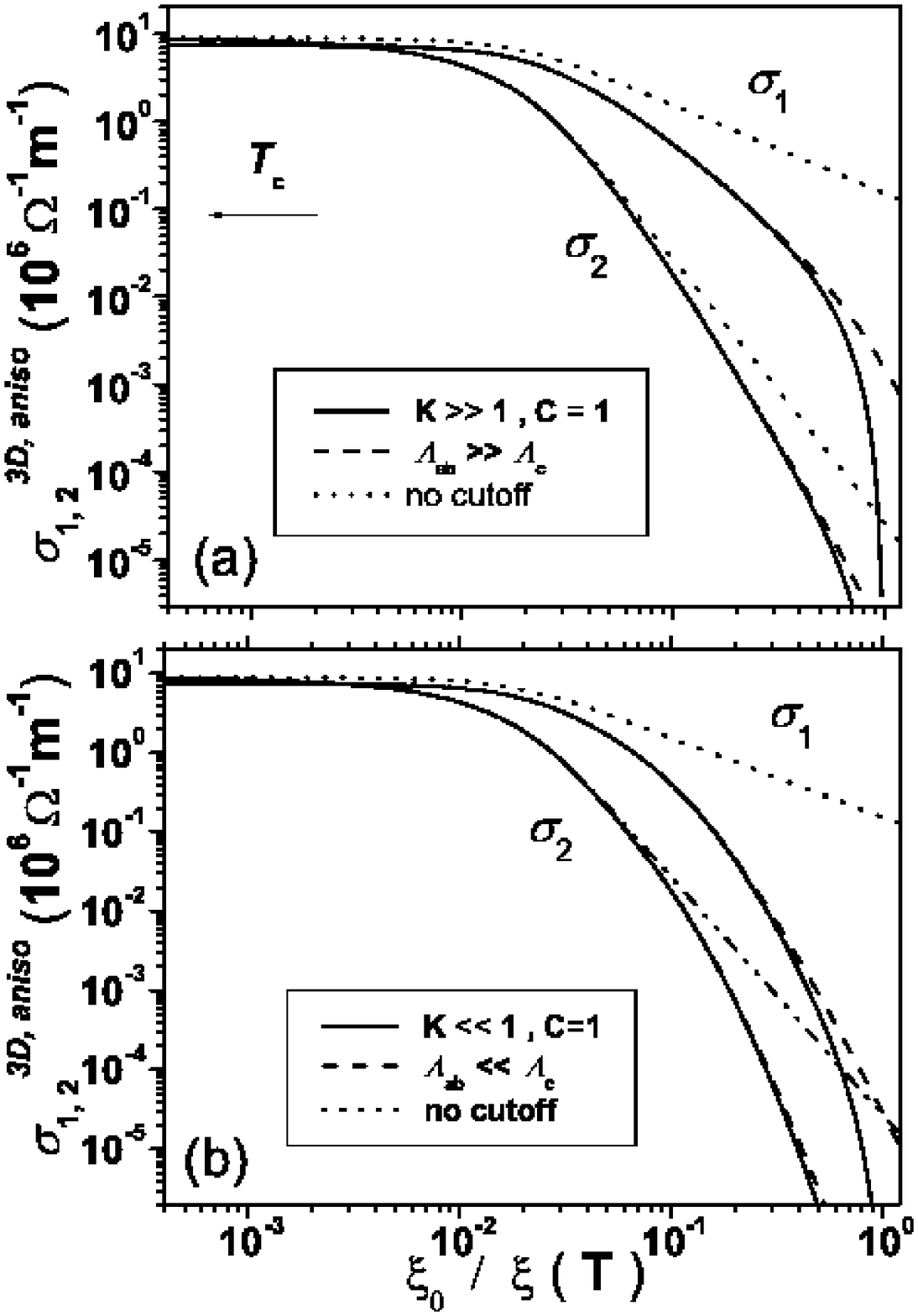}}
\centerline{\includegraphics[width=0.65\textwidth]{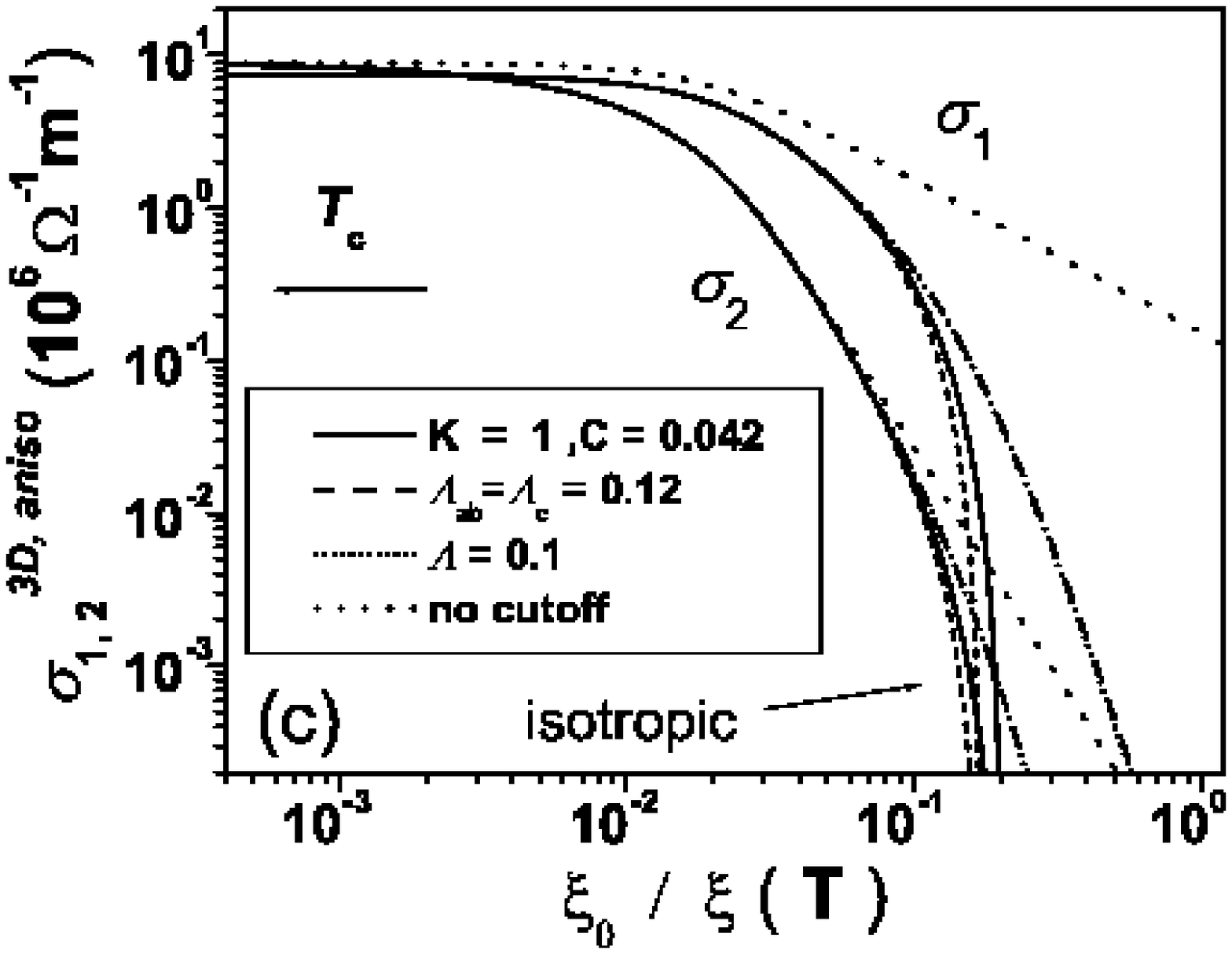}}
\caption{Calculated $\it ac$ fluctuation conductivities in the
anisotropic $\it 3D$ case. The ratio
$(\sigma_2(T_c)/\sigma_1(T_c)) = 1.2$ is assumed as in
Fig.~\ref{Fig1}. Three choices of the cutoff parameters are
selected: (a) $\Lambda_{ab} = 0.7$, $\Lambda_c = 0.06$, $C = 1$;
(b) $\Lambda_{ab}=0.1$, $\Lambda_c = 1$, $C =1 $; and
(c)$\Lambda_{ab} = \Lambda_c = 0.12$, $C = 0.042$, and $\Lambda=1$
for the isotropic case. The style of the presentation follows that
of Fig.~\ref{Fig1}.} \label{Fig3}
\end{figure}

If one assumes that the superconductor is anisotropic, one has to
deal with two cutoff parameters. Assuming the same value for the
ratio $(\sigma_2(T_c)/\sigma_1(T_c)) = 1.2$ as in Fig.~\ref{Fig1},
one can have the choices of the parameters $\Lambda_{ab}$ and
$\Lambda_c$ as shown in Fig.~\ref{Fig2} \cite{Peligrad:03}. An
alternative presentation is possible in terms of the ratio $K =
\Lambda_{ab}/\Lambda_c$ and the energy cutoff parameter $C$ as
shown in the inset to Fig.~\ref{Fig2}. The lower branch of this
curve corresponds to the cases $\Lambda_{ab} \ll \Lambda_c$ while
the upper branch is for $\Lambda_{ab} \gg \Lambda_c$. The
crossover $\Lambda_{ab} = \Lambda_c$ is shown by the dotted line
in the main panel of Fig.~\ref{Fig2}, and corresponds to $K = 1$
in the inset. In Fig.~\ref{Fig3}(a)-(c) we present the calculated
$\sigma$'s for three largely different choices of the cutoff
parameters. The style of the presentation follows that of
Fig.~\ref{Fig1}. We have argued earlier\cite{Peligrad:03} that
only the case with $\Lambda_{ab} \gg \Lambda_c$ is physically
acceptable in anisotropic superconductors where $\xi_{0ab} \gg
\xi_{0c}$. Here we present in Fig.~\ref{Fig3}(a) the results for
energy cutoff and compare them to those for wavevector cutoff. The
curves calculated with no cutoff are always added for the sake of
reference. One can observe that both, $\sigma_1$ and $\sigma_2$
have well pronounced slopes at higher temperatures. The effect of
the cutoff is to change those slopes. The effect of the energy
cutoff is observable only at very high temperatures where the
fluctuation conductivity rapidly vanishes. The choice
$\Lambda_{ab} \ll \Lambda_c$ for the wavevector cutoff, and the
accompanying choice $K \ll 1$ and $C = 1$ for the energy cutoff,
are shown in Fig.~\ref{Fig3}(b). The shape of these curves are
quite different from those of the previous case, thus showing the
sensitivity of the numerical calculations to the choices of the
cutoff parameters. Fortunately, the freedom of taking a diversity
of choices for the cutoff parameters is restricted by physical
arguments which make a large $\Lambda_c$ unacceptable in systems
having very small $\xi_{0c}$, as is the case in high-$T_c$
superconductors \cite{Peligrad:03}. Detailed analysis of the
experimental data presented below, provide further support of this
rule. Finally, we present also in Fig.~\ref{Fig3}(c) the case
$\Lambda_{ab}=\Lambda_c$ which appears to be very similar, but not
identical, to that of the isotropic case. It will also be shown as
inadequate for the experimental data analysis shown later in this
paper.
\par
\begin{figure}
\centerline{\includegraphics[width=0.8\textwidth]{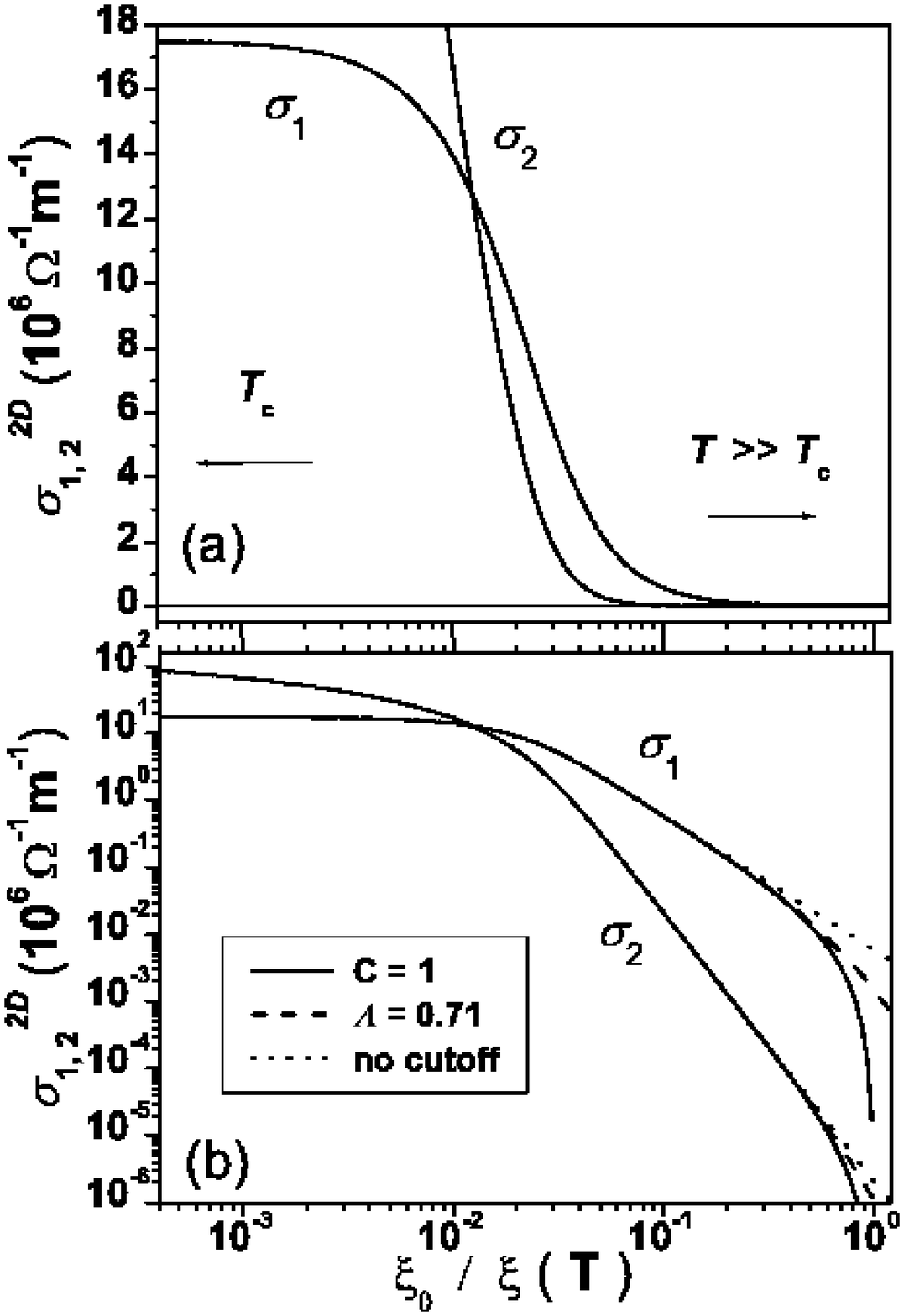}}
\caption{Calculated $\it ac$ fluctuation conductivities in the
$\it 2D$ case. For the sake of comparison, the $\it 2D$ cutoff
parameter ($\Lambda = 0.7$) is given the same value as
$\Lambda_{ab}$ of the $\it 3D$ case in Fig.~\ref{Fig3}(a). The
style of the presentation follows that of Fig.~\ref{Fig1}.}
\label{Fig4}
\end{figure}

It is of interest for the data analysis in this paper to examine
also the cutoff effects in two dimensional ($\it 2D$) cases. The
$\it ac$ fluctuation conductivity is then given by
\cite{Peligrad:03} %%
\begin{equation}
\displaystyle {\widetilde {\sigma}}^{\hspace{0.05cm}{\it 2D}} =
\frac{e^2}{16\hbar s}\left(\frac{\xi(T)}{\xi_0}\right)^z
\int_0^{Q}
\frac{4\, q^{\,3}\, \left[1 - i \Omega (1+q^2)^{-1}\right]}
{(1+q^2)[\Omega^2+(1+q^2)^2]}
\,d q \,\,\, ,
\label{4}
\end{equation}
where $s$ is the effective layer thickness. The wavevector cutoff
is made only in the $ab$-plane with
$k_{ab}^{max}=\sqrt{2}\Lambda/\xi_{0ab}$ while along the c-axis
only the lowest $k_c=0$ term is taken. Eq.~(\ref{3}) can again be
applied for the energy cutoff. The calculated curves for
$\sigma_1^{\it 2D}$ and $\sigma_2^{\it 2D}$ are shown in
Fig.~\ref{Fig4}. We have used the $\it 2D$ cutoff parameter
$\Lambda = 0.7$, i.~e. equal to the choice of $\Lambda_{ab}$ in
the $\it 3D$ case of Fig.~\ref{Fig3}(a). In contrast to the $\it
3D$ case, we observe that the cutoff does not bring about a change
in the slopes of $\sigma_1^{2D}$ and $\sigma_2^{2D}$ at higher
temperatures.
\par
\begin{figure}
\centerline{\includegraphics[width=1.2\textwidth]{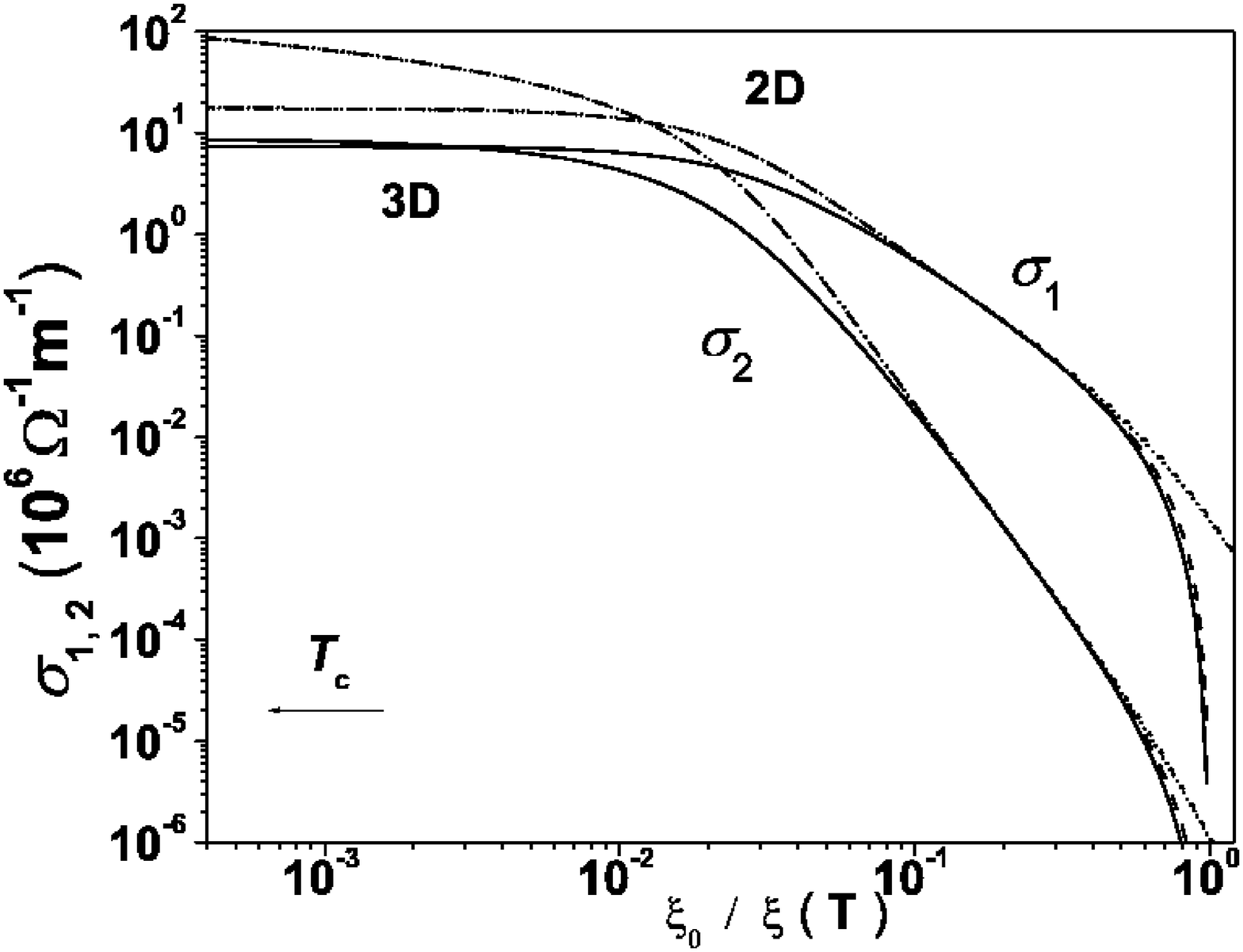}}
\caption{Direct comparison of the $\it 3D$ and $\it 2D$ curves
from Fig.~\ref{Fig3}(a) and Fig.~\ref{Fig4}, respectively,
calculated with cutoff.} \label{Fig5}
\end{figure}

The slopes of the $\sigma$'s at higher temperatures are believed to be
essential in an effort to determine the dimensionality of the fluctuations
in an experimental analysis. Indeed, the slopes of the curves calculated
with no cutoff in the $\it 3D$ and $\it 2D$ cases are different. However,
the cutoff increases the slopes in the $\it 3D$ case as shown in
Fig.~\ref{Fig3}(a). A direct comparison of the $\it 2D$ and
$\it 3D$ curves in Fig.~\ref{Fig5} shows that the cutoff makes the
distinction between the two cases more difficult. The parameters $\xi_{0c}$
and $s$ in the prefactors of Eq.~(\ref{1}) and Eq.~(\ref{4}) determine
just the offsets in the logarithmic plot of Fig.~\ref{Fig5}. These have been
chosen so as to match the values of $\it 2D$ and $\it 3D$ calculations at
higher temperatures. At temperatures closer to $T_c$ it is easy to determine
the dimensionality of the fluctuations, but at higher temperatures the two
cases become indistinguishable.
\par
\begin{figure}
\centerline{\includegraphics[width=0.65\textwidth]{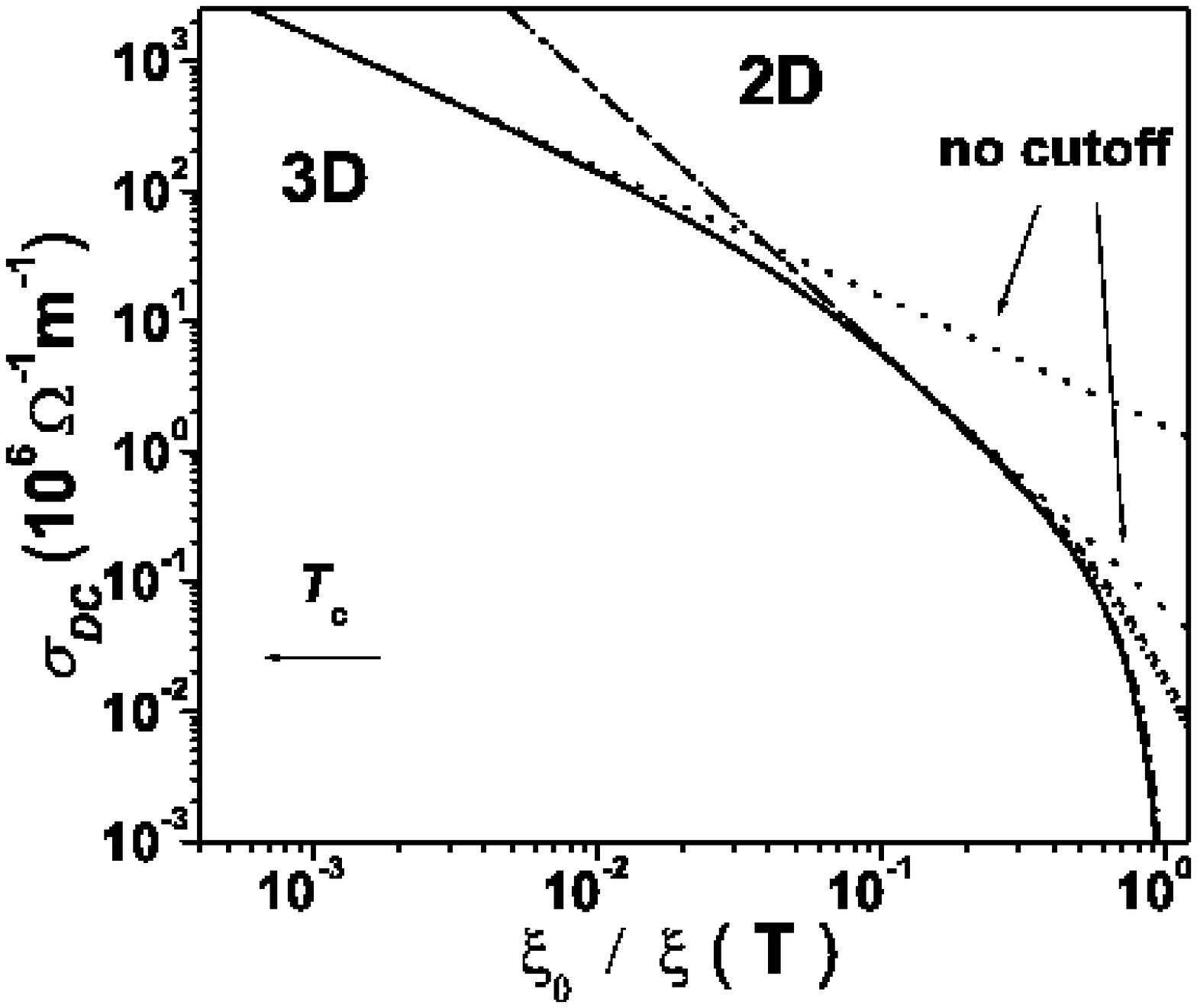}}
\caption{Calculated $\it dc$ fluctuation conductivities in the
$\it 3D$ (full line), and $\it 2D$ (dashed line) cases with cutoff
parameters as in Fig.~\ref{Fig3}(a) and Fig.~\ref{Fig4},
respectively. The calculations with no cutoff yield dotted lines
with slopes $-1/2$, and $-1$ for the $\it 3D$ and $\it 2D$ cases,
respectively.} \label{Fig6}
\end{figure}

It is worth noting that the above feature is pertinent not only to the
$\it ac$ but also to the $\it dc$ fluctuation conductivity.
Fig.~\ref{Fig6} demonstrates that the cutoff affects strongly
$\sigma_{dc}^{\it 3D}$ beyond some temperature, so that it becomes
indistinguishable from that of $\sigma_{dc}^{\it 2D}$.
Hence, the dimensionality of the fluctuations cannot be infered
from the behavior of the $dc$ fluctuation conductivity at higher
temperatures, provided that the cutoff effects are properly accounted for.
\par
We may also remark that if only the wavevector cutoff were
considered, the curves in Fig.~\ref{Fig6} would acquire slope $-6$
in the limit of very high temperatures. If the coherence length
acquired the Gaussian form $(\xi(T)/\xi_0)=1/\sqrt{\epsilon}$ at
those temperatures, the $dc$ fluctuation conductivity would have
the behavior $\sigma_{dc} \propto 1/\epsilon^3$ in both, $\it 2D$
and $\it 3D$ cases \cite{Cimberle:97,Silva:01,Silva:02}. However,
significant deviations from this behavior have been observed
experimentally, and ascribed to the energy cutoff
\cite{Carballeira:01,Vina:02}.
\par
\begin{figure}
\centerline{\includegraphics[width=0.65\textwidth]{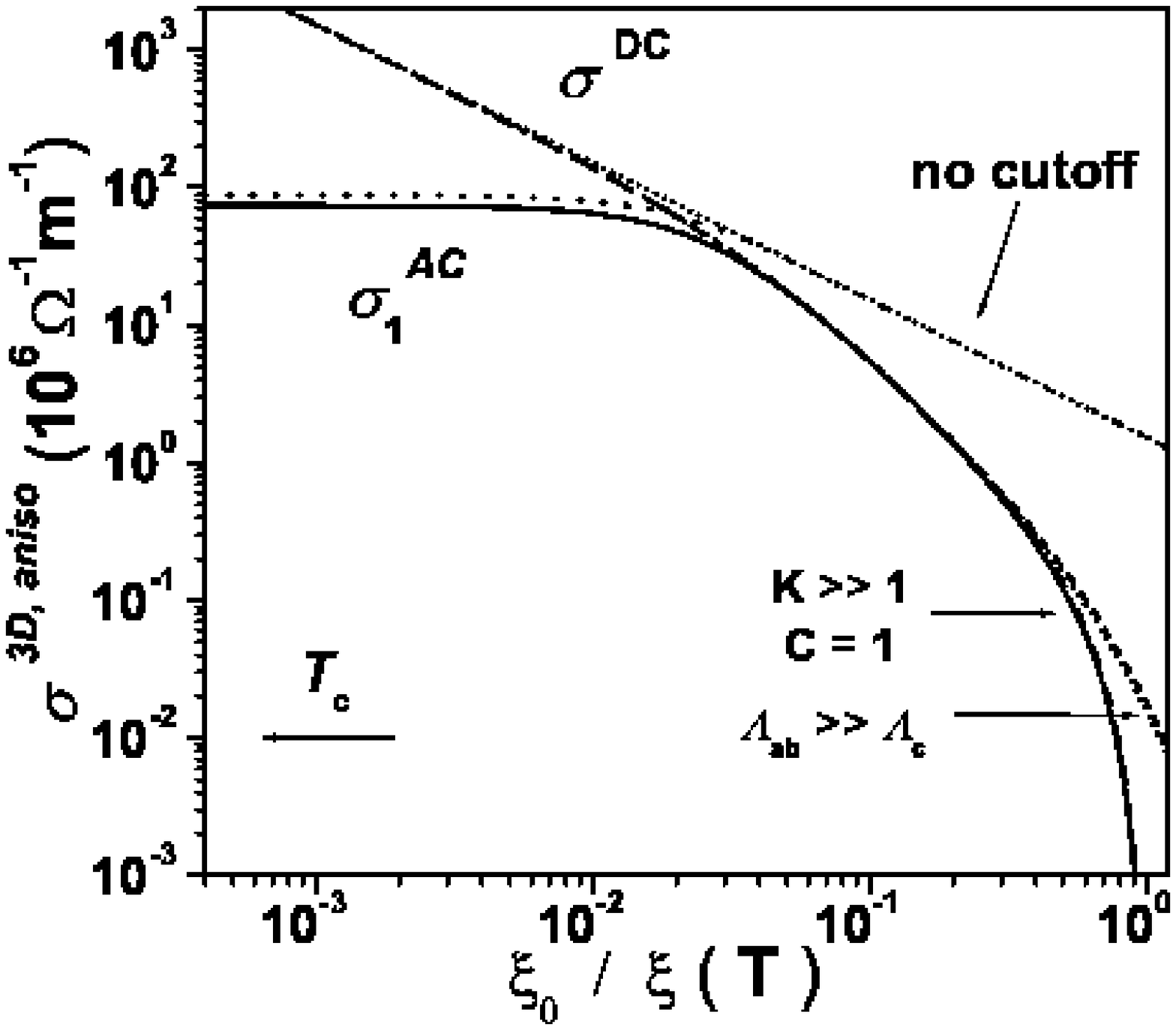}}
\caption{Comparison of the $\it dc$ fluctuation conductivity and
the real part $\sigma_1$ of the $\it ac$ case calculated with the
same set of parameters in the $\it 3D$ expressions.} \label{Fig7}
\end{figure}

Finally, it may be of interest to compare the behavior of the $\it
dc$ fluctuation conductivity and the real part of the fluctuation
conductivity in the $\it ac$ case. Fig.~\ref{Fig7} shows the
corresponding $\it 3D$ curves calculated with the same set of
parameters. Obviously, the fluctuation conductivity in the $\it
dc$ case diverges when $T_c$ is approached, while $\sigma_1$ of
the $\it ac$ case reaches a finite value. However, at higher
temperatures, the information gained from a $\it dc$ measurement
is equivalent to that of the real part $\sigma_1$ in the $\it ac$
case \cite{Peligrad:03}. The imaginary part $\sigma_2$ is an
additional information provided by an $\it ac$ measurement.
\par
Note that here we have considered only the fluctuation conductivity, but,
for the sake of simplicity, no subscript has been used . The total
conductivity above $T_c$ includes also a contribution from the normal
electrons. The latter has to be added to both, $\sigma_{dc}$ and the real
part $\sigma_1$ of the $\it ac$ fluctuation conductivity, but not to the
imaginary part $\sigma_2$ which is due solely to the superconducting
electrons.
\section{EXPERIMENTAL} \label{Sect3}
We have measured a number of YBCO and BSCCO thin films grown on
various substrates. The main features in our results did not
change with the change of the substrate or thickness of the film.
Here we report specifically on the measurements in an YBCO thin
film (thickness 200 nm) grown on $MgO$ substrate, BSCCO-2212 (350
nm) on $LaAlO_3$ and BSCCO-2223 (100 nm) on $NdGaO_3$. The sample
was cut typically to 4 mm in length and 1 mm in width and mounted
on a sapphire sample holder which extends to the center of an
elliptical copper cavity resonating in $_{\rm e}$TE$_{111}$ mode
at $\approx 9.5$~GHz. The sample was oriented with its longest
side along the microwave electric field $E_{\omega}$. In this
configuration the microwave current flows only in the $ab$-plane
of the film. The sapphire sample holder was thermally connected to
a heater and sensor assembly but isolated from the body of the
microwave cavity. With a temperature controller the temperature of
the sample could be varied from 2 K up to room temperature.
However, the cavity was kept in pumped helium flow at 1.7 K in
order to eliminate spurious signals from cavity heating.
\par
The measured quantity is the complex frequency shift $\Delta
\widetilde{\omega}/\omega = \Delta f/f + i\,\Delta (1/2Q)$ in
which the resonant frequency $f$ and the $Q$-factor of the cavity
change with the sample temperature. The empty cavity had $1/2Q$
close to 20 ppm, which was subtracted from the data measured with
the sample. The level of $1/2Q$ with the sample in the normal
state could be several hundred parts per million (ppm) and we were
interested in detecting small changes due to the superconducting
fluctuations above $T_c$. For this purpose it was beneficial to
use the recently introduced modulation technique,
\cite{Nebendahl:01} which enables the resolution of $\Delta(1/2Q)$
to 0.02 ppm. The resonant frequency of the cavity was measured by
a microwave frequency counter.
\par
For the film in the microwave electric field the cavity perturbation
analysis yields \cite{Peligrad:01}
\begin{equation}
\displaystyle \frac{\Delta \widetilde{\omega}}{\omega}=
\frac{\Gamma}{N}\left[1-N+\frac{(\widetilde{k}/k_0)^2 N}
{\left[\coth(i\widetilde{k}d/2)
+\tanh(i\widetilde{k}\zeta)\right]i\widetilde{k}d/2}\right]^{-1} \, ,
\label{5}
\end{equation}
where $\Gamma$ is the filling factor of the sample in the cavity, and
$N$ is the depolarization factor of the film. The intrinsic property
of the sample is its complex conductivity
$\widetilde{\sigma} = \sigma_1 - i \sigma_2$. It is contained in the complex
wavevector $\displaystyle \widetilde{k} =
k_0 \sqrt{1 -i \widetilde{\sigma}/(\epsilon_0\omega)}$,
where $\displaystyle k_0=\omega\sqrt{\mu_0\epsilon_0}$ is the
vacuum wavevector. The thickness of the film
is $d$, and $\zeta$ is the asymmetry parameter due to the
substrate.\cite{Peligrad:01}  The unknown parameters in
Eq.~(\ref{5}) can be evaluated from the ratio of the slopes of the
experimental curves $\Delta (1/2Q)$ and $\Delta f/f$ in the normal
state far above $T_c$, provided that the normal state conductivity
is known at the temperature where the experimental
slopes were evaluated. Eq.~(\ref{5}) can then be used to convert the
experimental data for $\Delta (1/2Q)$ and $\Delta f/f$ at any
temperature to obtain the corresponding experimental values of
$\sigma_1$ and $\sigma_2$.
\section{BSCCO-2212} \label{Sect4}
\begin{figure}
\centerline{\includegraphics[width=0.8\textwidth]{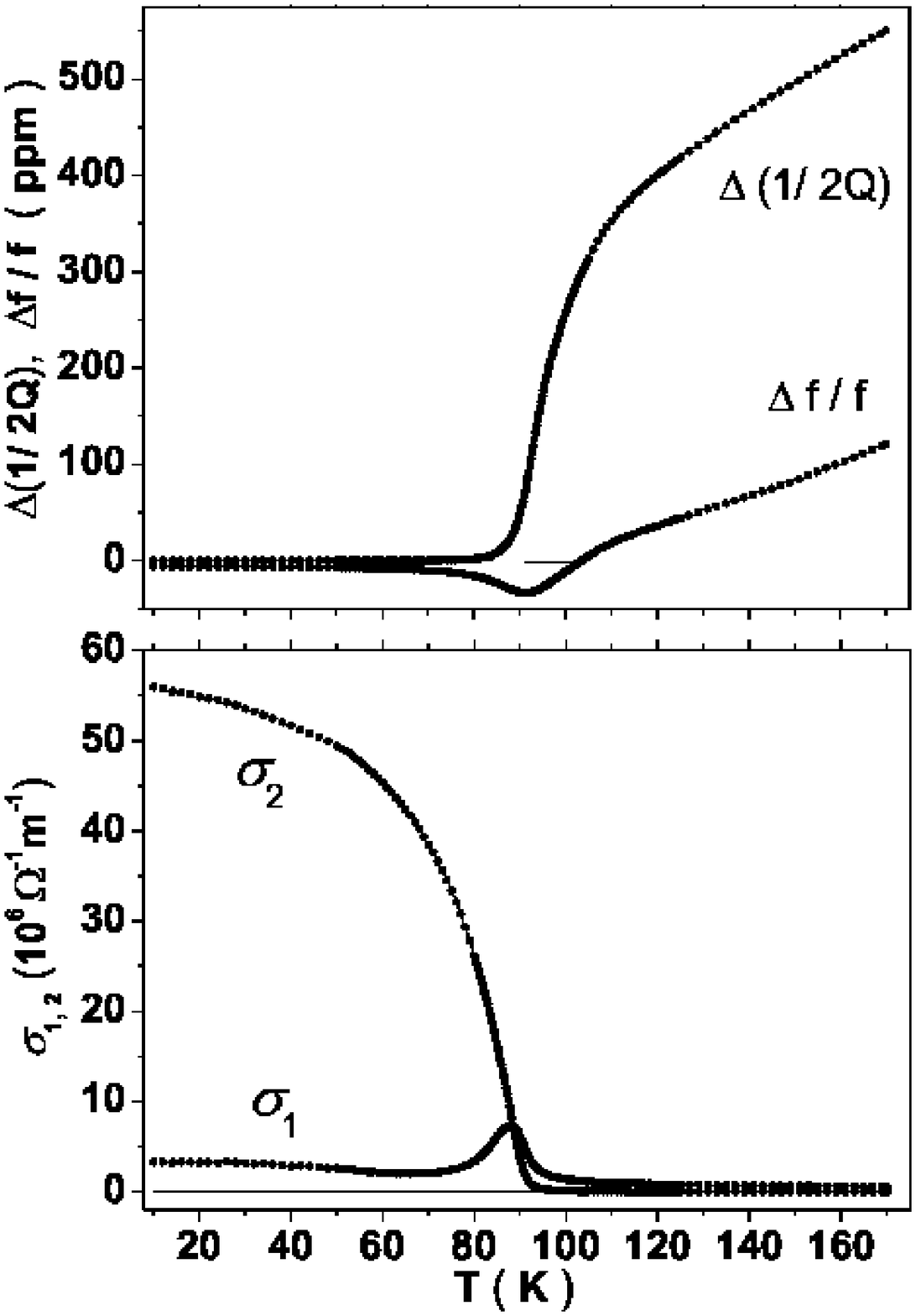}}
\caption{Measured complex frequency shift in BSCCO-2212 thin film
(a), and the deduced complex conductivity (b).} \label{Fig8}
\end{figure}

Fig.~\ref{Fig8} shows the experimental data of BSCCO-2212 thin
film and the complex conductivity deduced from these data by means
of Eq.~(\ref{5}). One can observe clearly the peak in the real
part ($\sigma_1$) due to the superconducting fluctuations. Its
maximum appears when the coherence length diverges,
\cite{Dorsey:91,Fisher:91} and this occurs at the critical
temperature of the superconducting transition. The imaginary part
($\sigma_2$) rises from its zero value found above $T_c$, to a
saturation at low temperatures.
\par
The total experimentally determined conductivity above $T_c$ may
include the contributions due to the normal conductivity and the
superconducting fluctuation conductivity. In all previously
reported $\it dc$ conductivity measurements, it was assumed that
the experimentally observed conductivity was the sum
$\sigma^{exp}=\sigma_n + \sigma$. The normal conductivity near
$T_c$ was obtained from the extrapolated linear resistivity far
above $T_c$. Hence, the fluctuation conductivity $\sigma$ was
obtained upon subtraction $\sigma^{exp} - \sigma_n$, and then
analyzed. This approach, however, neglects the effect of the
pseudogap which has been observed by other techniques in
high-$T_c$ superconductors \cite{Timusk:99}. The opening of the
pseudogap reduces the density of one electron states at the Fermi
level so that the normal conductivity is also reduced below
$\sigma_n$. The pseudogap is observed in optimally doped samples
near $T_c$, and in underdoped samples even well above $T_c$. If
the pseudogap is fully opened near $T_c$, the normal conductivity
should completely vanish, and the whole experimentally observed
conductivity should be due to the fluctuation conductivity, i.~e.
$\sigma^{exp} = \sigma$.
\par
\begin{figure}
\centerline{\includegraphics[width=0.8\textwidth]{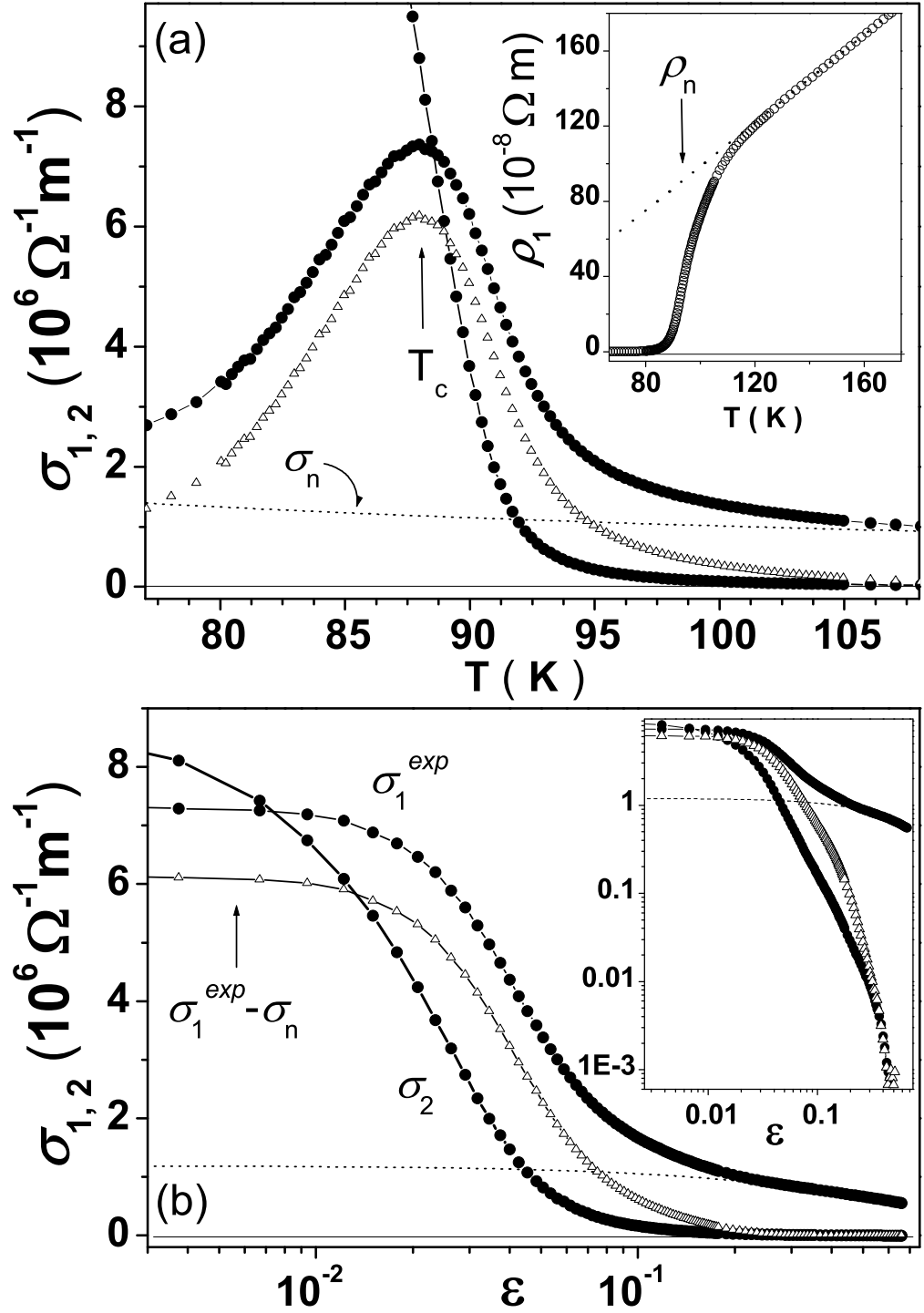}}
\caption{(a) The real part of the fluctuation conductivity in
BSCCO-2212 taken as either the total experimental value
$\sigma_1=\sigma_1^{exp}$ ($\bullet$), or as the value obtained
upon subtraction $\sigma_1 = \sigma_1^{exp}-\sigma_n$
($\triangle$), where $\sigma_n$ is the normal conductivity
obtained by extrapolation of the linear resistivity far above
$T_c$ (inset). The imaginary part $\sigma_2$ is background free
and needs no subtraction. (b) Fluctuation conductivities above
$T_c$ as functions of the reduced temperature $\epsilon =
\ln{\left(T / T_c\right)}$.} \label{Fig9}
\end{figure}

We have analyzed our measured microwave conductivity, with and
without subtracting the extrapolated $\sigma_n$ from the real part
$\sigma_1^{exp}$. Fig.~\ref{Fig9}(a) shows on an enlarged scale
around $T_c$ the two sets of data, i.~e. $\sigma_1^{exp}$, and
$\sigma_1^{exp}-\sigma_n$. The inset in Fig.~\ref{Fig9}(a) shows a
linear $\rho_n$ fitted to the real part of the resistivity at
temperatures above 150~K. This linear $\rho_n$ is shown
extrapolated to lower temperatures. It is used to calculate the
extrapolated normal conductivity $\sigma_n = 1/\rho_n$ shown by
the dotted line in the main panel of Fig.~\ref{Fig9}(a). Note that
the imaginary part $\sigma_2$ is due solely to the superconducting
fluctuations, i.~e. $\sigma_2$ is background free so that no
subtraction should be in place. It is obvious that the
experimental ratio $\sigma_2/\sigma_1$ at $T_c$ is different in
the two sets of data. This ratio is important in the determination
of the cutoff parameters as discussed in Section~\ref{Sect2}. In
this paper, we study the conductivities above $T_c$ shown in
Fig.~\ref{Fig9}(b) as functions of the reduced temperature
$\epsilon = \ln{\left(T / T_c\right)}$.
\par
\begin{figure}
\centerline{\includegraphics[width=0.8\textwidth]{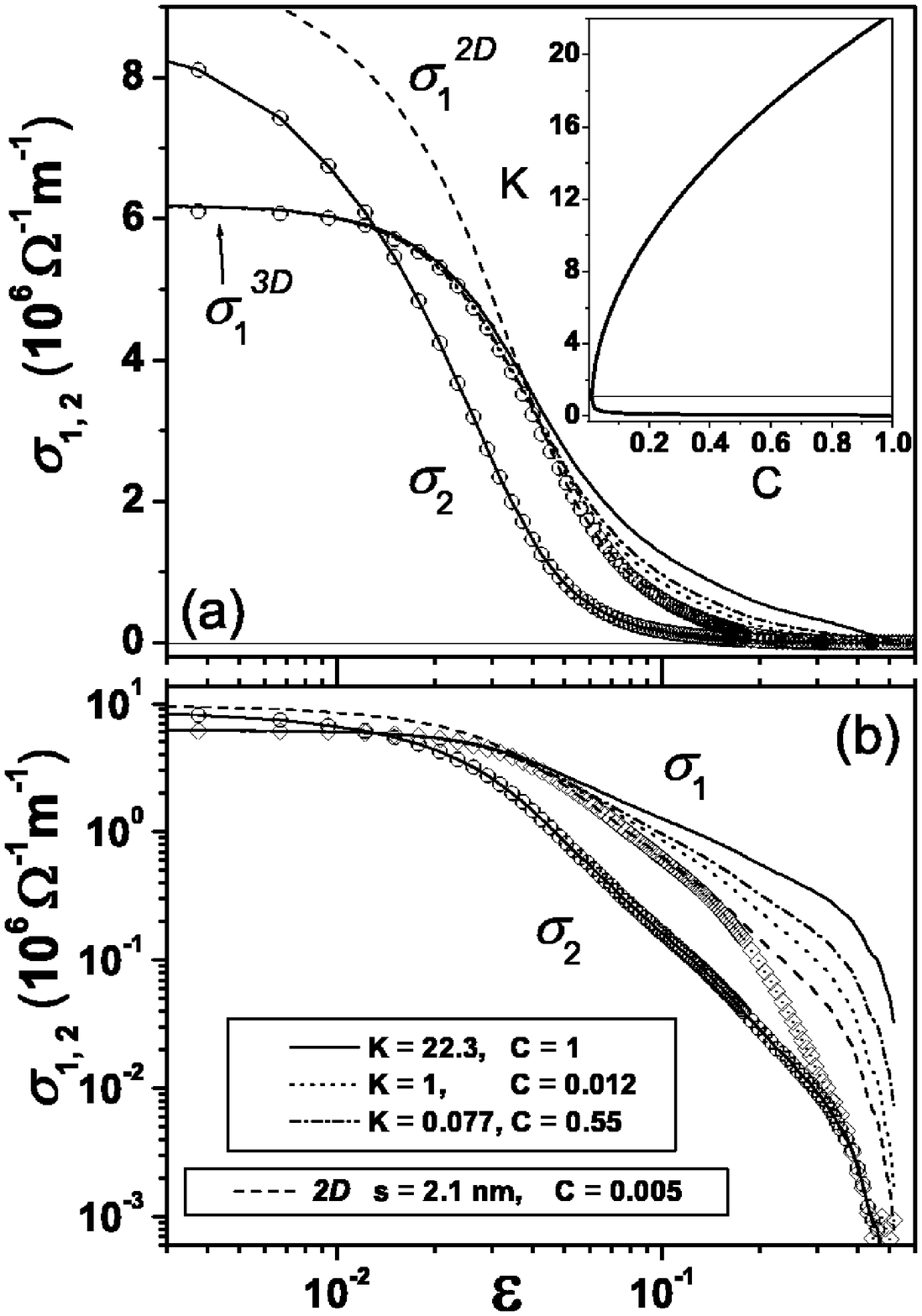}}
\caption{The real part $\sigma_1 = \sigma_1^{exp}-\sigma_n$ and
the imaginary part $\sigma_2$ of the fluctuation conductivity
above $T_c$ in BSCCO-2212. The inset shows the possible choices of
the cutoff parameters for $\sigma_2/\sigma_1$ = 1.46 at $T_c$. The
results of the calculations using the anisotropic $\it 3D$ and
$\it 2D$ expressions are presented by the various curves in the
main panel. The concomitant coherence lengths are given in
Fig.~\ref{Fig11}.} \label{Fig10}
\end{figure}
\begin{figure}
\centerline{\includegraphics[width=0.8\textwidth]{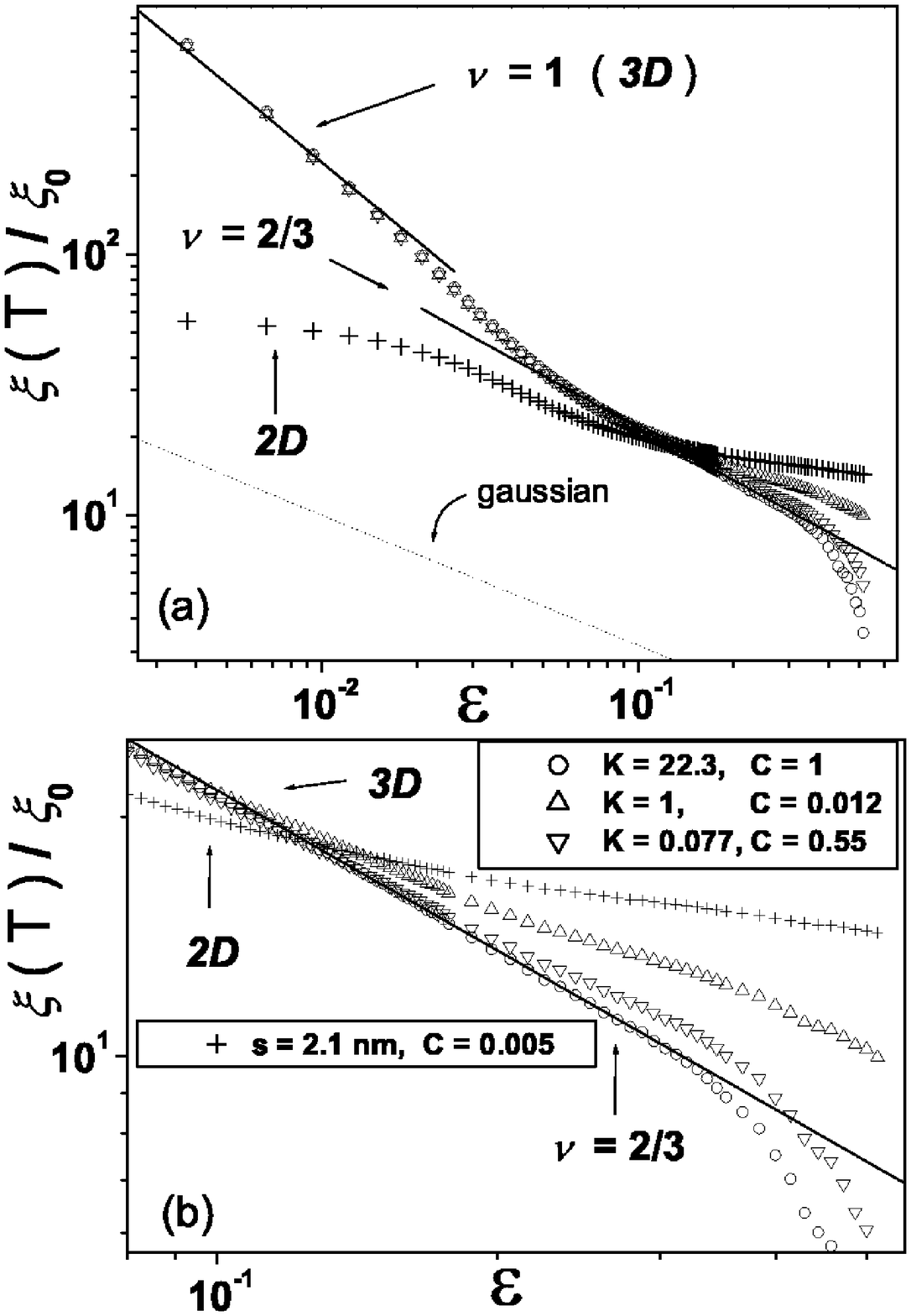}}
\caption{The coherence lengths calculated from $\sigma_2$ using
the various expressions and cutoff parameters as indicated in
Fig.~\ref{Fig10}. The lower panel shows on an enlarged scale the
slope 2/3 (full line) and the behavior of the various cases at
higher temperatures.} \label{Fig11}
\end{figure}

We shall first consider the conventional approach in which
$\sigma_n$ is subtracted. In that case we find
$\sigma_2/\sigma_1$=1.46 at $T_c$, and use this value to set a
constraint on the choices of the cutoff
parameters.\cite{Peligrad:03} The inset in Fig.~\ref{Fig10}(a)
shows the relationship between the parameters $K =
\Lambda_{ab}/\Lambda_c$ at $T_c$ and the energy cutoff parameter
$C$ defined in Eq.~(\ref{3}). Taken an allowed choice of the
cutoff parameters, one can equate the imaginary part of
Eq.~(\ref{1}) to the experimental value of $\sigma_2$, and solve
numerically for the reduced coherence length $\xi(T)/\xi_0$. Note
that the parameter $\xi_{0c}$ in the prefactor of Eq.~(\ref{1})
can be determined straightforwardly from the value of $\sigma_2$
at $T_c$ where $\xi(T)/\xi_0$ diverges. Namely, the value of
$\sigma_2$ at $T_c$ is practically insensitive to the choice of
the cutoff parameters which can, for that purpose, be set to any
value \cite{Peligrad:03}. Here we obtained $\xi_{0c}=0.05$~nm.
Once the values of the reduced coherence length $\xi(T)/\xi_0$ are
calculated for all the experimental points, one can use them in
the real part of Eq.~(\ref{1}) to calculate $\sigma_1$ and compare
with the experimental data. Fig.~\ref{Fig10} shows the results for
several choices of the cutoff parameters. The concomitant
coherence lengths are presented in Fig.~\ref{Fig11}. Near $T_c$
one obtains the same values for the coherence length regardless of
the choice of the cutoff parameters. This is due to the fact that
$\sigma_2$ is practically insensitive to the wavevector cutoff in
that temperature region. It appears that very close to $T_c$ the
coherence length follows the critical exponent $\nu = 1$. The
differences between the various choices appear at higher
temperatures.
\par
Let us first look at the anisotropic $\it 3D$ case with $C = 1$
and $K \gg 1$. It yields the coherence length in Fig.~\ref{Fig11}
having the critical exponent $\nu = 2/3$ pertaining to the $3D$ XY
universality class. This behavior would be in agreement with the
results of other experiments
~\cite{Pasler:98,Meingast:01,Osborn:02}. However, the calculated
$\sigma_1$ fits the experimental data only near $T_c$ as seen in
Fig.~\ref{Fig10}. At higher temperatures the calculated $\sigma_1$
overestimates the experimental fluctuation conductivity
represented by the data set $\sigma_1 = \sigma_1^{exp}-\sigma_n$.
This overestimation has no physical explanation. Therefore, we
should reject this choice of the cutoff parameters.
\par
The choice $K = 1$ entails the condition $C \ll 1$ so that the energy cutoff
is more severe. The calculated $\sigma_1$ is therefore more reduced at
higher temperatures, and the overall fit to the experimental points appears
to be improved (cf. Fig.~\ref{Fig10}). However, some overestimation is still
present, in particular at high temperatures, so that this solution should
also be rejected. Even stronger argument for the rejection comes
from the inspection of the concomitant coherence length in Fig.~\ref{Fig11}.
At higher temperatures, its slope is reduced well below the
value 2/3 found in other experiments. This is particularly well seen
in Fig.~\ref{Fig11}(b) where high temperature region is presented on an
enlarged scale. The reduced slope appears as the mathematical
result of the strong energy cutoff $C \ll 1$.
\par
For $K \ll 1$ the calculated $\sigma_1$ overestimates largely the
experimental values at higher temperatures. Hence, this choice of the cutoff
parameters can be rejected, too.
\par
We may also look at the results of the $\it 2D$ calculation based on the
expression given in Eq.~(\ref{4}). One can adjust the parameters so that the
calculated $\sigma_1$ fits the experimental data very well at higher
temperatures as seen in Fig.~\ref{Fig10}. Therefore, one might be tempted to
consider the scenario in which the system obeys the anisotropic $\it 3D$
behavior with $K \gg 1$ and $C = 1$ at temperatures close to $T_c$ followed
by a crossover to a $\it 2D$ behavior at higher temperatures. This scenario
could yield an overall good fit to the experimental data, except at very
high temperatures (cf. Fig.~\ref{Fig10}(b)). However, the coherence length
obtained in the $\it 2D$ calculation is physically unacceptable because of
the saturation at high temperatures as seen in Fig.~\ref{Fig11}.
Having explored all the possibilities, we have to conclude that none
of the above scenarios yields a satisfactory result with the data set
$\sigma_1 = \sigma_1^{exp}-\sigma_n$.
\par
\begin{figure}
\centerline{\includegraphics[width=0.8\textwidth]{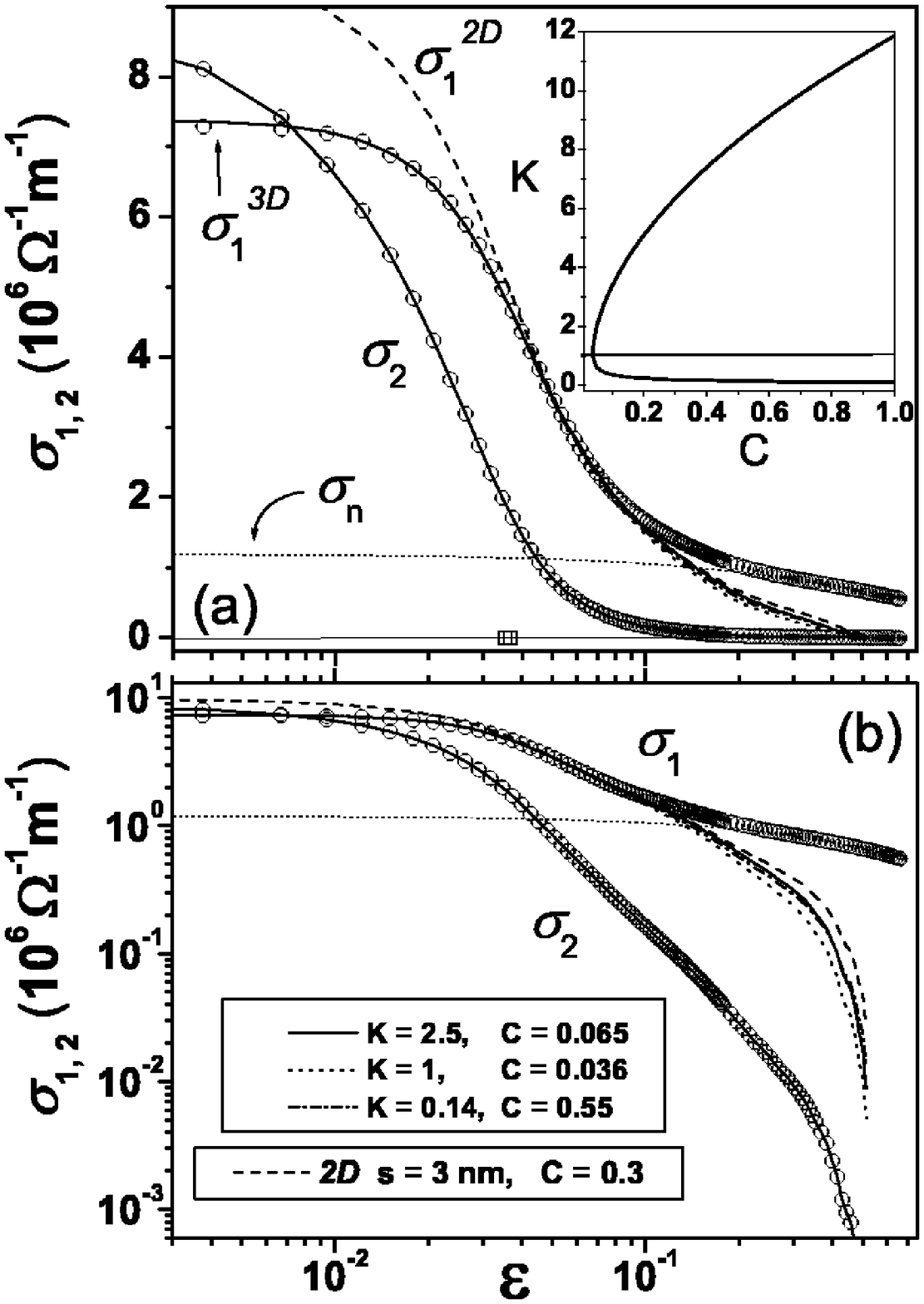}}
\caption{The fluctuation conductivity in BSCCO-2212 taken as
$\sigma_1 = \sigma_1^{exp}$ and $\sigma_2$ as measured. The inset
shows the possible choices of the cutoff parameters for
$\sigma_2/\sigma_1=1.22$ at $T_c$. The calculations were made for
the $\it 3D$ and $\it 2D$ cases as indicated in the legend and
described in the text. The concomitant coherence lengths are
presented in Fig.~\ref{Fig13}.} \label{Fig12}
\end{figure}
\begin{figure}
\centerline{\includegraphics[width=0.8\textwidth]{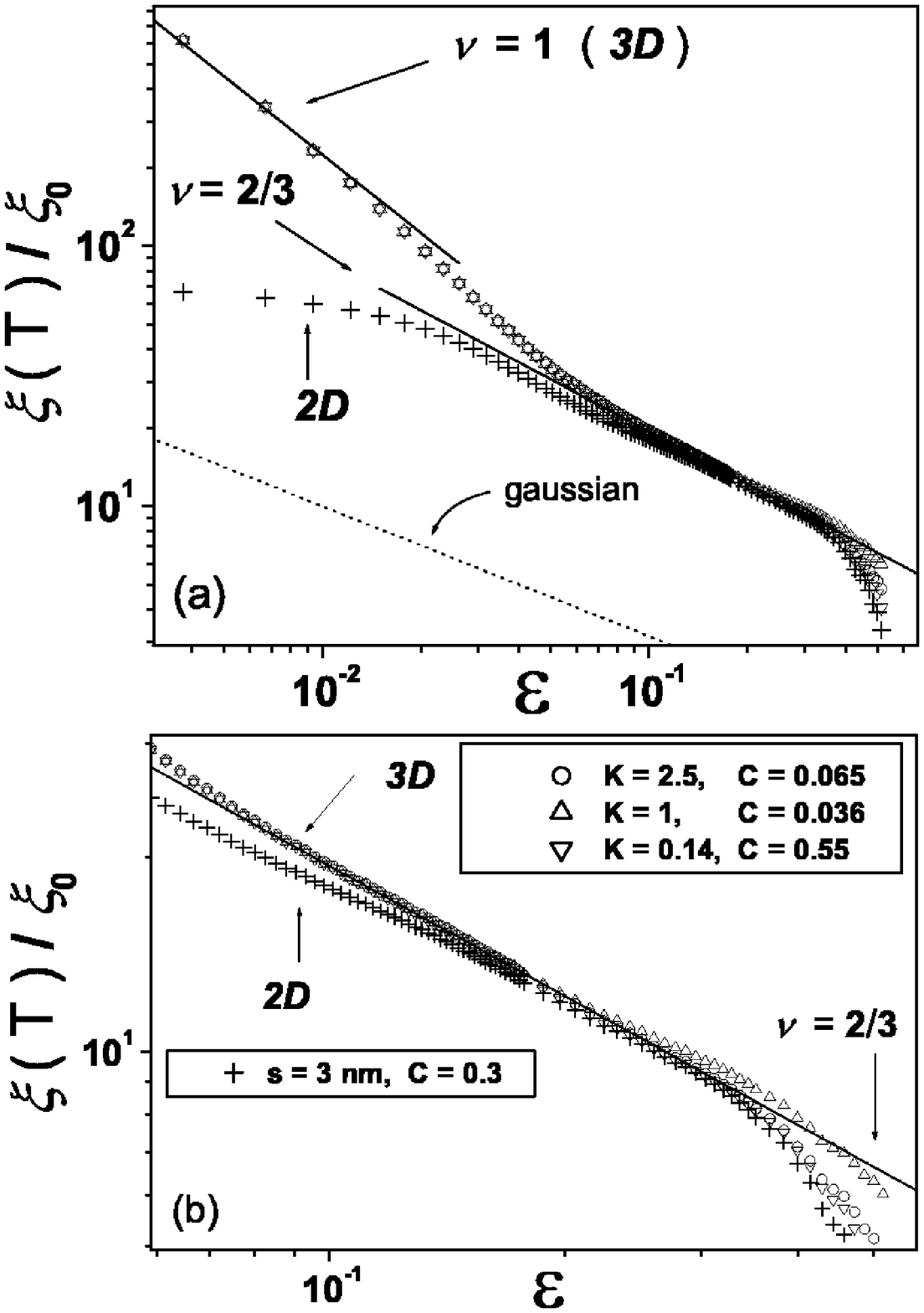}}
\caption{The coherence lengths calculated from $\sigma_2$ in the
various cases indicated in Fig.~\ref{Fig12}. The lower panel shows
the high temperature behavior on an enlarged scale.} \label{Fig13}
\end{figure}

Next, we turn to the case in which no subtraction is made so that the whole
experimental $\sigma_1^{exp}$ near $T_c$ is attributed to the fluctuation
conductivity $\sigma_1$. From Fig.~\ref{Fig9} we find first
$\sigma_2/\sigma_1=1.22$ at $T_c$, and find the corresponding curve for the
cutoff parameters in the inset of Fig.~\ref{Fig12}(a). It is quite different
from that of the preceding case in the inset of Fig.~\ref{Fig10}(a).
The data analysis can be done as before and the results for the various
choices of the cutoff parameters are shown in Fig.~\ref{Fig12}, while the
concomitant coherence lengths are presented in Fig.~\ref{Fig13}.
\par
The highest temperature where $\sigma_2$ is still detectable
beyond the noise level is $\epsilon = 0.5$ (145~K).
At that temperature, $\sigma_2$ has decreased by four orders of magnitude
from its value at $T_c$ (Fig.~\ref{Fig12}(b)), and further on our signal
is lost in the experimental noise. Our method of analysis relies
on the experimental values of $\sigma_2$ to calculate the reduced coherence
length. Hence, the latter is also available only up to this same temperature
(Fig.~\ref{Fig13}). The same applies to $\sigma_1$ which is calculated from
the coherence length. Beyond this temperature, the measured conductivity
contains only the nonzero real part $\sigma_1^{exp}$, which
is shown in Fig.~\ref{Fig12} up to $\epsilon = 0.66$ (170~K).
\par
One observes in Fig.~\ref{Fig12} that the anisotropic $\it 3D$ expressions
with $K=2.5$ and $C=0.065$ yield $\sigma_1$ which matches very well the
experimental data up to $\epsilon \approx 0.1$. At higher temperatures the
calculated values deviate from the experimental points, but in this case the
deviation falls below the data and can be physically explained. Namely, the
calculated $\sigma_1$ is the fluctuation contribution. At temperatures close
to $T_c$ there is no normal conductivity contribution if the pseudogap is
fully opened so that the calculated fluctuation conductivity $\sigma_1$
equals the measured $\sigma_1^{exp}$. Beyond some temperature, the pseudogap
starts to close and the normal conductivity contribution appears gradually
growing from zero to its full value $\sigma_n$ at high temperatures. This
growing normal conductivity contribution is seen in Fig.~\ref{Fig12} as the
difference between the experimental points and the calculated fluctuation
conductivity $\sigma_1$ shown by the full line. A more detailed discussion
of the pseudogap will be given in a subsequent section.
\par
The other choices of the cutoff parameters used in the $\it 3D$ calculations
are also presented in Fig.~\ref{Fig12}. The fits of the calculated $\sigma_1$
to the experimental data are less good. Also, the coherence length does not
follow the expected slope 2/3 in Fig.~\ref{Fig13}.
\par
The $\it 2D$ analysis is particularly interesting. One can choose the
parameters so that the calculated $\sigma_1$ is practically indistinguishable
from that of the best anisotropic $\it 3D$ choice in Fig.~\ref{Fig12}.
Also, the coherence length obtained in the $\it 2D$ calculation approaches
the slope 2/3 at higher temperatures. The indistinguishability of the
$\it 3D$ and $\it 2D$ behavior at higher temperatures has already been
discussed in Section \ref{Sect2}. Here we may examine whether a $\it 3D$
to $\it 2D$ crossover is a possible scenario. A crude criterion for the
crossover is that $\xi_c(\epsilon)$ becomes comparable to $s/2$. The
$\it 2D$ curves in Fig.~\ref{Fig12} were calculated with $s=3$~nm so that
the above criterion is met when $\xi_c(\epsilon^*) \approx 1.5$~nm.
Since $\xi_{0c} = 0.05$~nm is determined from Eq.~(\ref{1}) and the
experimental value of $\sigma_2$ at $T_c$, one finds
$\xi_c(\epsilon^*)/\xi_{0c} \approx 30$. From Fig.~\ref{Fig13} one finds
that this condition corresponds to $\epsilon^* \approx 0.06$. This value
is in accord with the observation of a good fit of the calculated
$\sigma_1^{\it 2D}$ to the experimental data in Fig.~\ref{Fig12}. From that
point of view, the $\it 3D$-$\it 2D$ crossover appears as a likely scenario.
However, the coherence length in the $\it 2D$ case does not reach the
slope 2/3 at that temperature, but only at a much higher one
($\epsilon \approx 0.18$). Therefore, one does not find decisive arguments
in favor of the $\it 3D$-$\it 2D$ crossover in BSCCO-2212.
\par
Having established the consistent analysis of the data, we may
comment on the resulting coherence length. Fig.~\ref{Fig12}
reveals the existence of two critical regimes with the static
critical exponents $\nu = 2/3$ well above $T_c$, and a crossover
to $\nu = 1$ when $T_c$ is approached. The former critical regime
corresponds to the $3D$ XY universality class, and has been
reported before ~\cite{Pasler:98,Meingast:01,Osborn:02}. However,
the well defined crossover to the critical regime with $\nu = 1$
is novel and surprising. We may also remark that our analysis
yields not only the slopes, but also the absolute values of the
reduced coherence length. Thus, we observe that these absolute
values are very high, much higher than those predicted by the mean
field or Gaussian expression $(\xi(T)/\xi_0) = 1/\sqrt{\epsilon}$.
Note also that the crossover between the two critical regimes is
not just a change of the slopes but involves a step in the
absolute values of the reduced coherence length. We find that the
critical behaviour persists as far above $T_c$ as the reduced
coherence length is still large.
\par
Also important is to comment on the smallness of the parameter
$\xi_{0c} = 0.05$~nm found above. Such small values were also
obtained in the analysis of the $\it dc$ fluctuation conductivity
\cite{Hopfengartner:91}. It is to be noted that this parameter is
not the zero temperature coherence length, but the coefficient in
the linear term of the Ginzburg-Landau functional. The zero
temperature coherence length should have a larger, physically
acceptable dimension. Also, the obtained value for $\xi_{0c}$ is
certainly too small to represent the shortest length for the
variation of the order parameter. In fact, the above analysis
based on the energy cutoff yields min($\xi_c(T)$) = $\xi_{0c} /
\sqrt{C}$. In the present case with $C = 0.065$ as in
Fig.~\ref{Fig13}(a), one obtains min($\xi_c$) = 0.2 ~nm. This is
large enough to be physically acceptable for the shortest
variation of the coherence length along the c-axis.
\par
\begin{figure}
\centerline{\includegraphics[width=0.8\textwidth]{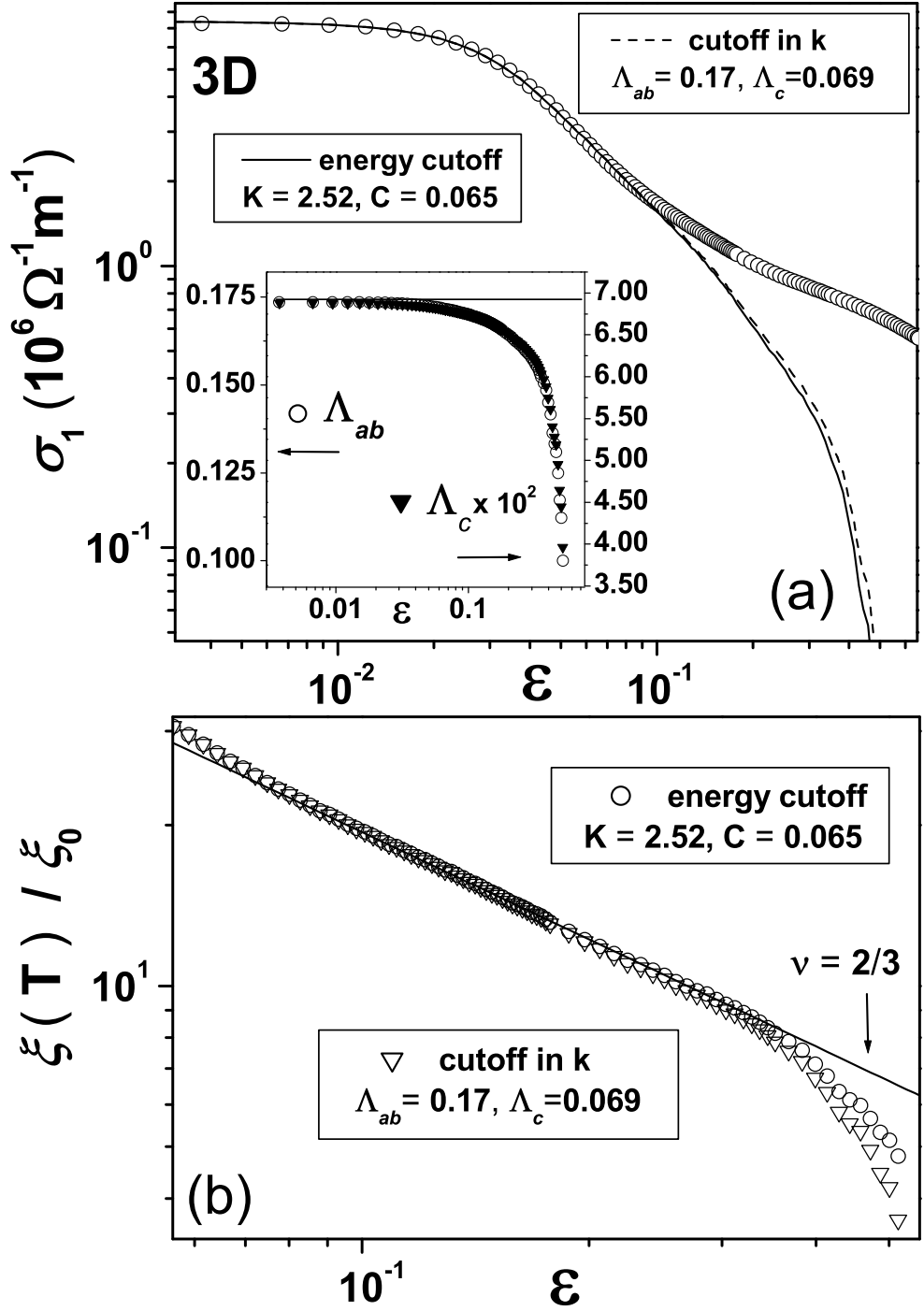}}
\caption{Direct comparison of the calculations made with the
wavevector cutoff and with the energy cutoff. In the former
approach the $\Lambda$'s are fixed at all temperatures while in
the latter these parameters are reduced at higher temperatures as
shown in the inset.} \label{Fig14}
\end{figure}

The above calculations were all carried out with energy cutoff, but the
main conclusions of the analysis remain valid even if only wavevector
cutoff is taken. Fig.~\ref{Fig14} shows a comparison of the calculations
with wavevector and energy cutoffs. In the former approach, $\Lambda_{ab}$
and $\Lambda_c$ are determined at $T_c$ by the experimental ratio
$\sigma_2(T_c)/\sigma_1(T_c)$, and then kept fixed at all temperatures.
On the contrary, in the energy cutoff approach, the values of $\Lambda_{ab}$
and $\Lambda_c$ are reduced at higher temperatures according to
Eq.~(\ref{3}). Their temperature variation is shown in the inset of
Fig.~\ref{Fig14}(a). One can readily observe that there is practically no
difference between the two approaches over most of the temperature range
covered by the actual data. Only at the highest temperatures one observes
a slightly stronger decrease of the calculated values when the energy
cutoff is applied.
\section{BSCCO-2223} \label{Sect5}
\begin{figure}
\centerline{\includegraphics[width=0.8\textwidth]{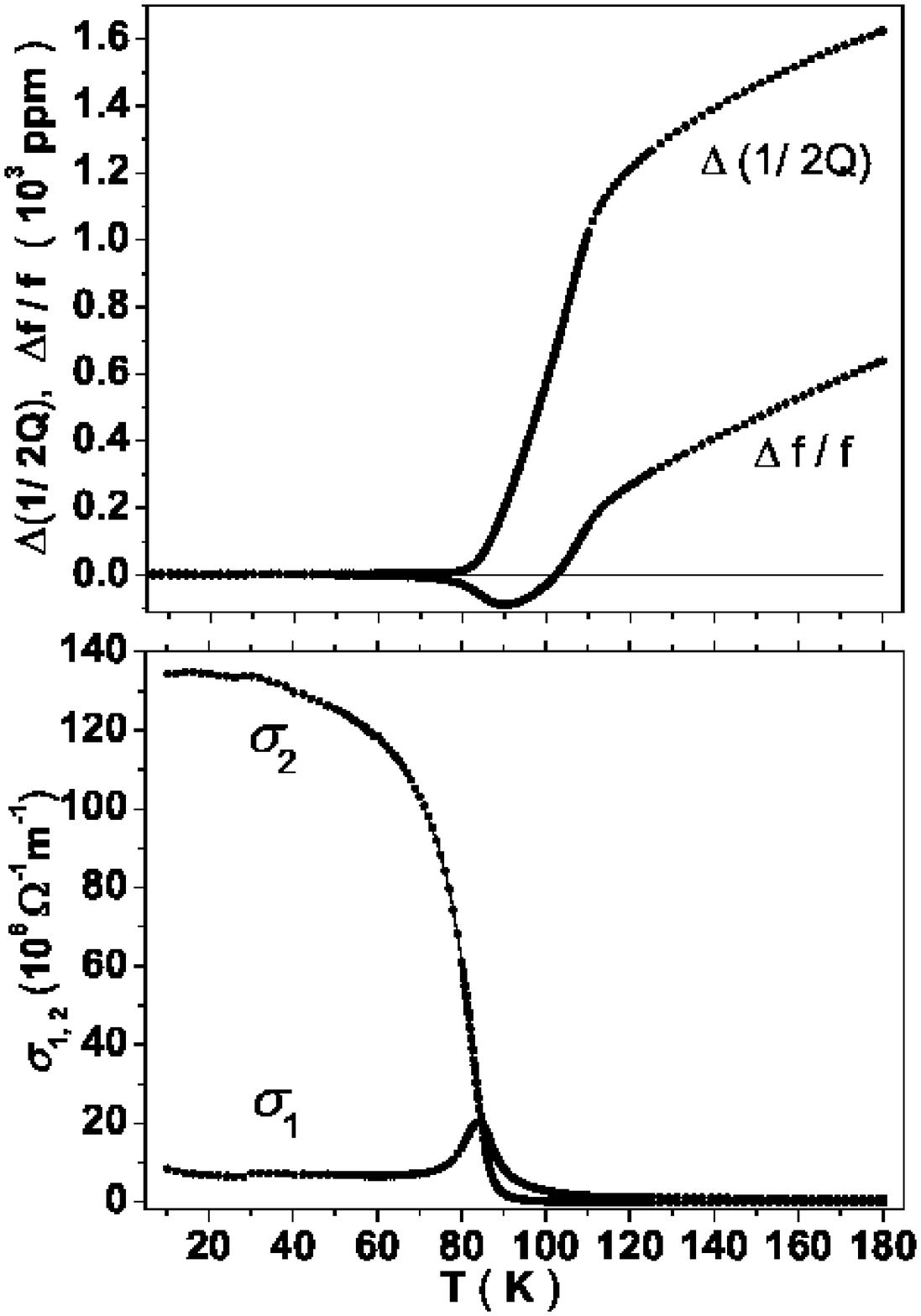}}
\caption{Measured complex frequency shift and conductivity in
BSCCO-2223 thin film.} \label{Fig15}
\end{figure}

\begin{figure}
\centerline{\includegraphics[width=0.8\textwidth]{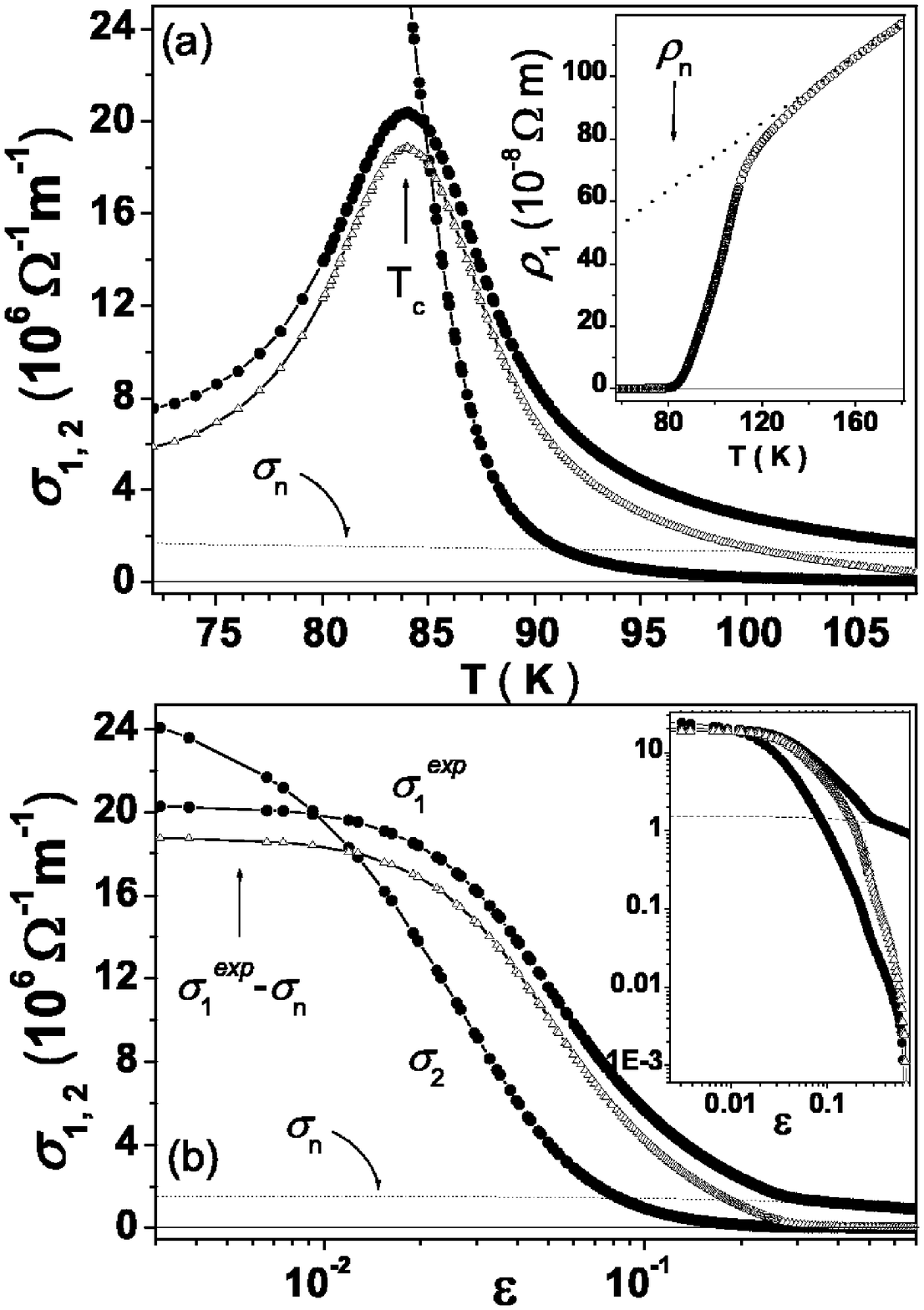}}
\caption{(a) Complex conductivity in BSCCO-2223 near $T_c$. The
extrapolated linear resistivity $\rho_n$ is shown in the inset.
The as measured $\sigma_1^{exp}$ ($\bullet$) and subtracted
$\sigma_1^{exp}-\sigma_n$ ($\triangle$) data sets are shown. (b)
Fluctuation conductivities above $T_c$ as functions of the reduced
temperature $\epsilon = \ln{\left(T / T_c\right)}$.} \label{Fig16}
\end{figure}

\begin{figure}
\centerline{\includegraphics[width=0.8\textwidth]{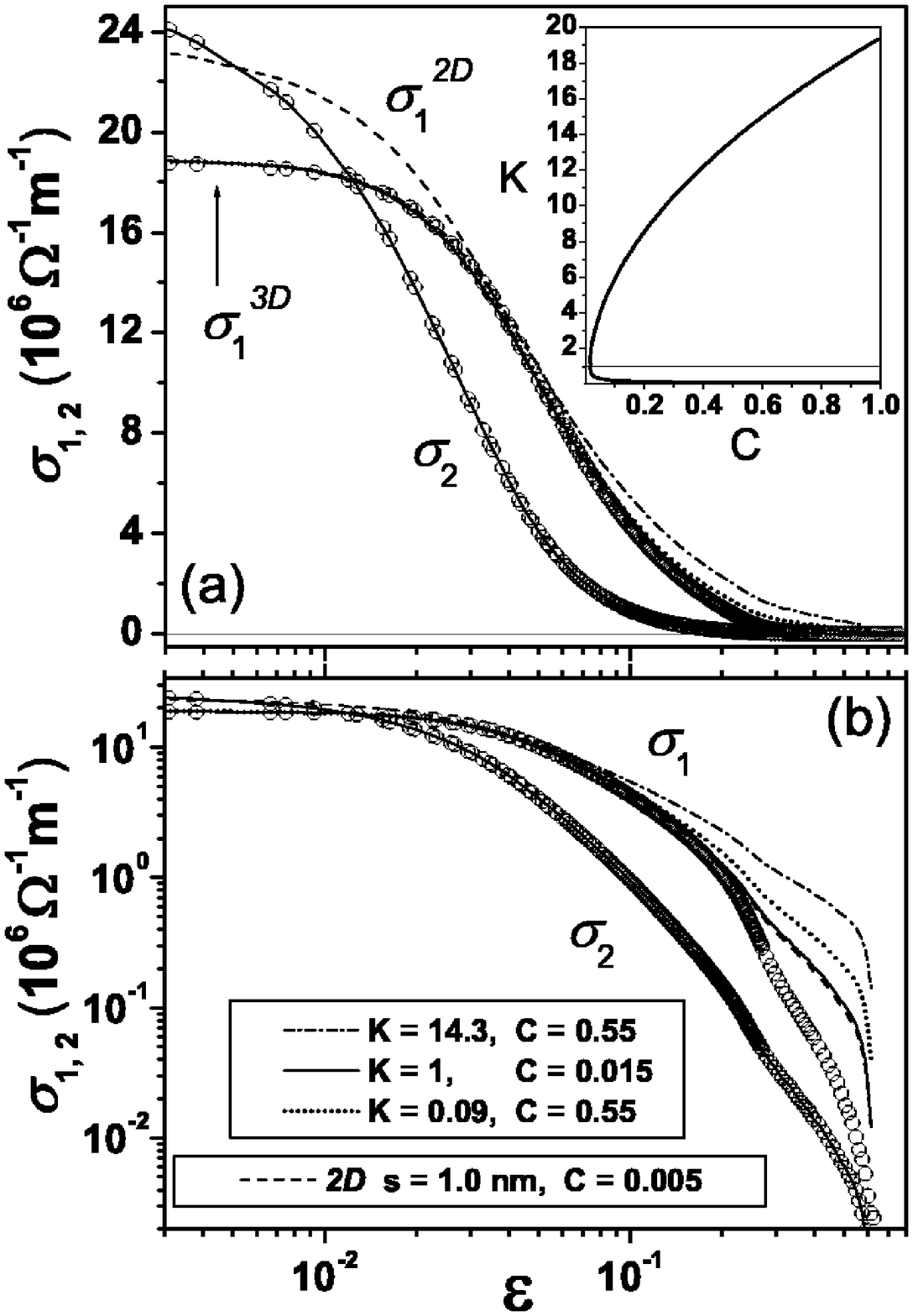}}
\caption{Data analysis on the subtracted data set
$\sigma_1^{exp}-\sigma_n$ and $\sigma_2$ in BSCCO-2223 thin film.
The presentation is parallel to that of Fig.~\ref{Fig10}. The
inset shows the possible choices of the cutoff parameters for
$\sigma_2/\sigma_1 = 1.4$ at $T_c$.} \label{Fig17}
\end{figure}

\begin{figure}
\centerline{\includegraphics[width=0.8\textwidth]{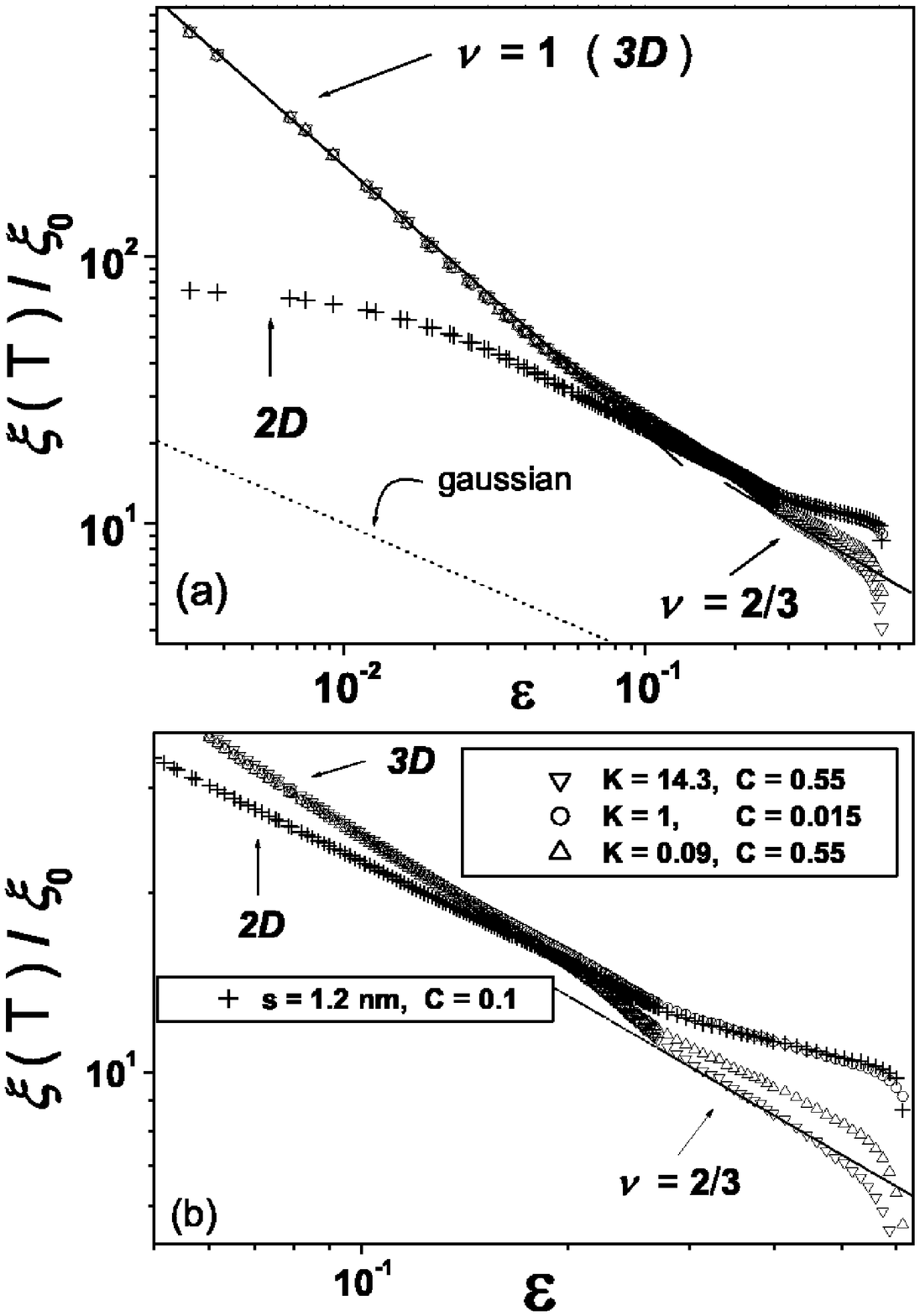}}
\caption{The reduced coherence length calculated from $\sigma_2$
in the cases presented in Fig.~\ref{Fig17}.} \label{Fig18}
\end{figure}

Fig.~\ref{Fig15} shows the measured complex frequency shift and the
conductivity in BSCCO-2223 thin film. The main features are similar to those
observed in BSCCO-2212 thin film. The transition here appears to be broader
and $T_c$ is lower, both indicating that the sample is underdoped. We have
carried out the complete analysis for each of the various cases as described
in the preceding section. Fig.~\ref{Fig16} shows the conductivities with and
without the subtraction of the extrapolated normal conductivity $\sigma_n$.
The analysis based on the subtracted data set is shown in Fig.~\ref{Fig17}
and the concomitant coherence lengths are shown in Fig.~\ref{Fig18}. One
can easily verify that none of the choices for the cutoff parameters can
yield satisfactory results with this data set. This is parallel to the
conclusion reached above in the case of BSCCO-2212.
\par
\begin{figure}
\centerline{\includegraphics[width=0.65\textwidth]{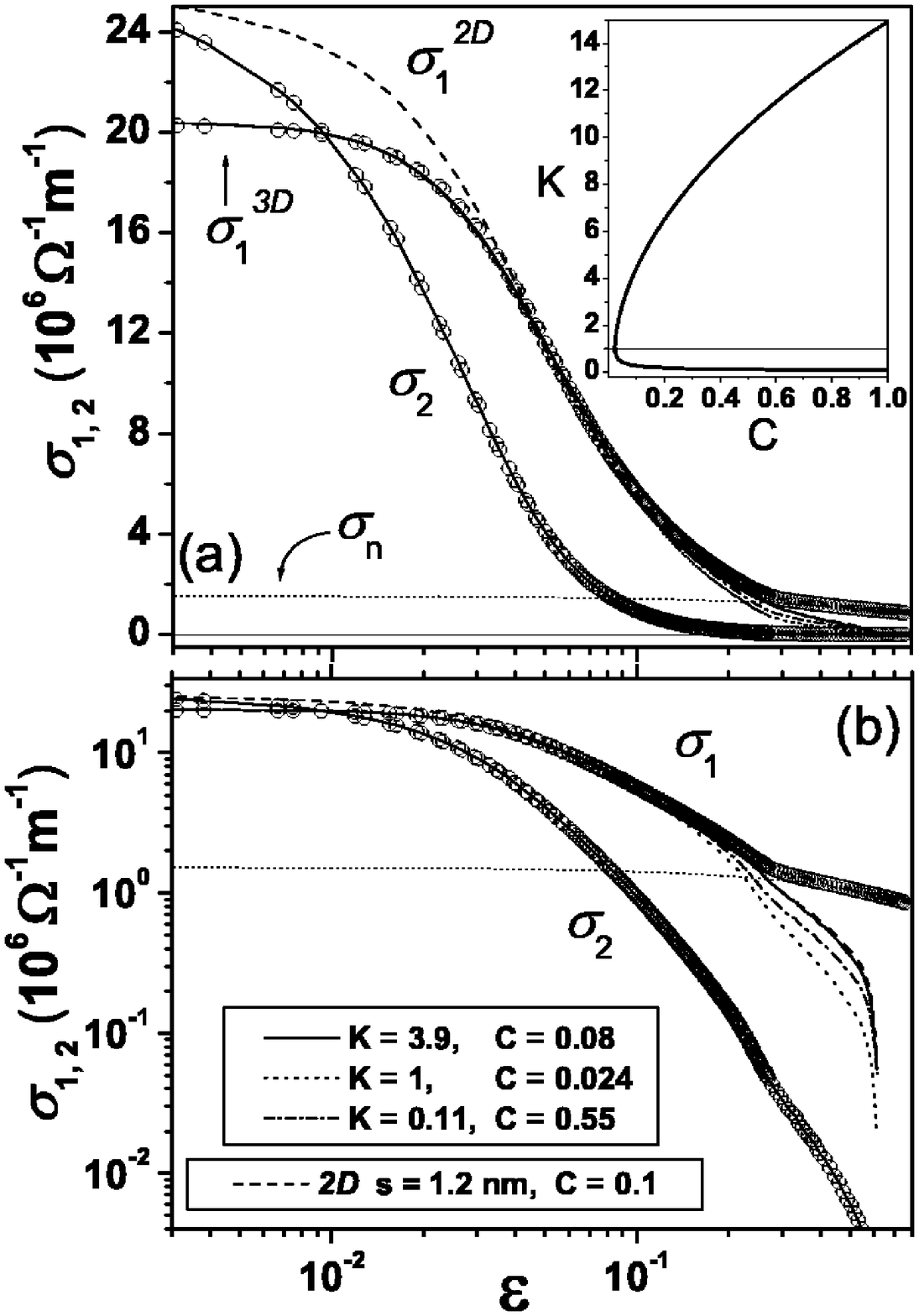}}
\centerline{\includegraphics[width=0.55\textwidth]{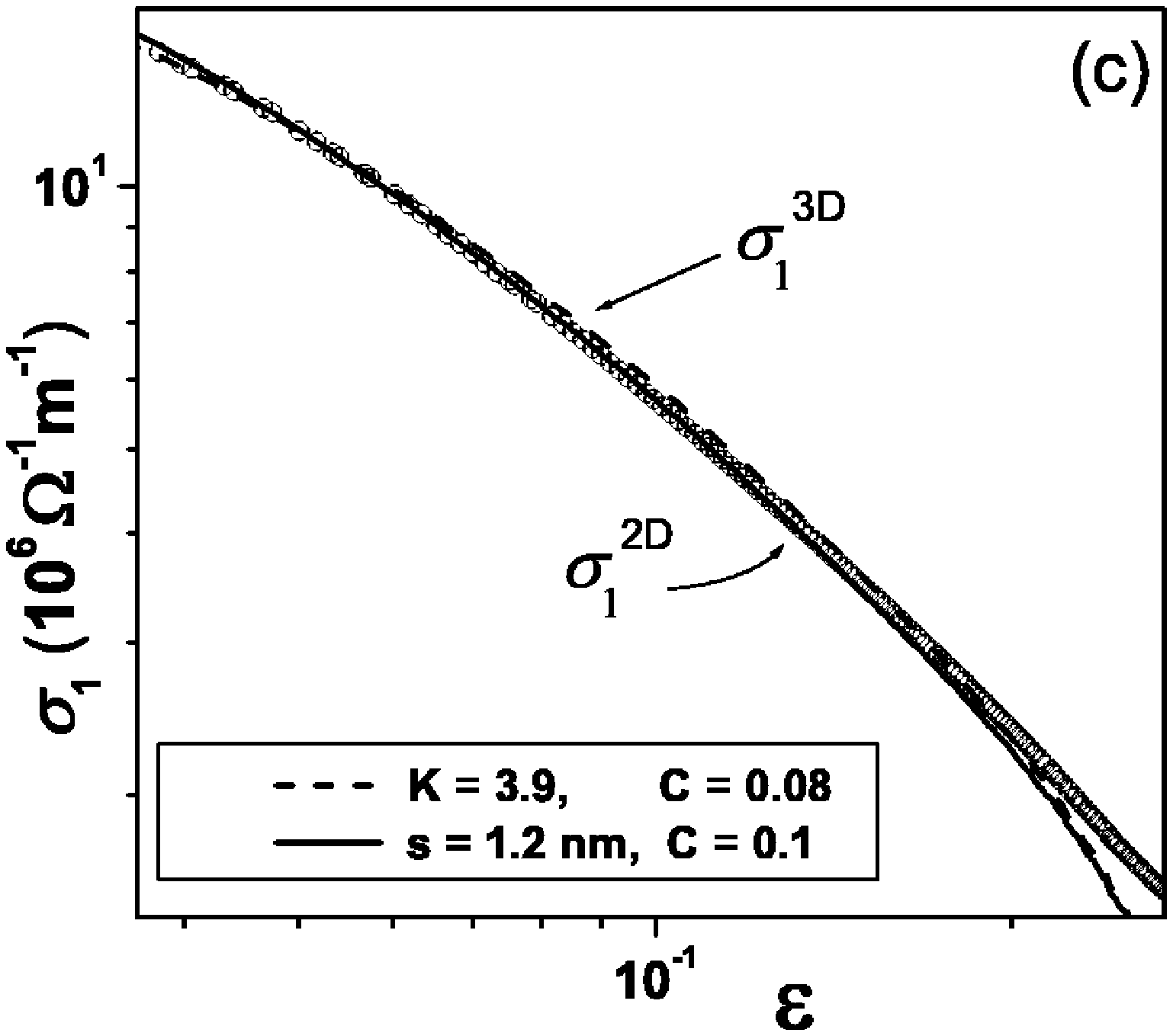}}
\caption{The fluctuation conductivity in BSCCO-2223 taken as
$\sigma_1 = \sigma_1^{exp}$ and $\sigma_2$ as measured. The inset
shows the possible choices of the cutoff parameters for
$\sigma_2/\sigma_1 = 1.29$ at $T_c$. The calculations were made
for the $\it 3D$ and $\it 2D$ cases as indicated in the legend and
described in the text. The concomitant coherence lengths are
presented in Fig.~\ref{Fig20}.} \label{Fig19}
\end{figure}
\begin{figure}
\centerline{\includegraphics[width=0.8\textwidth]{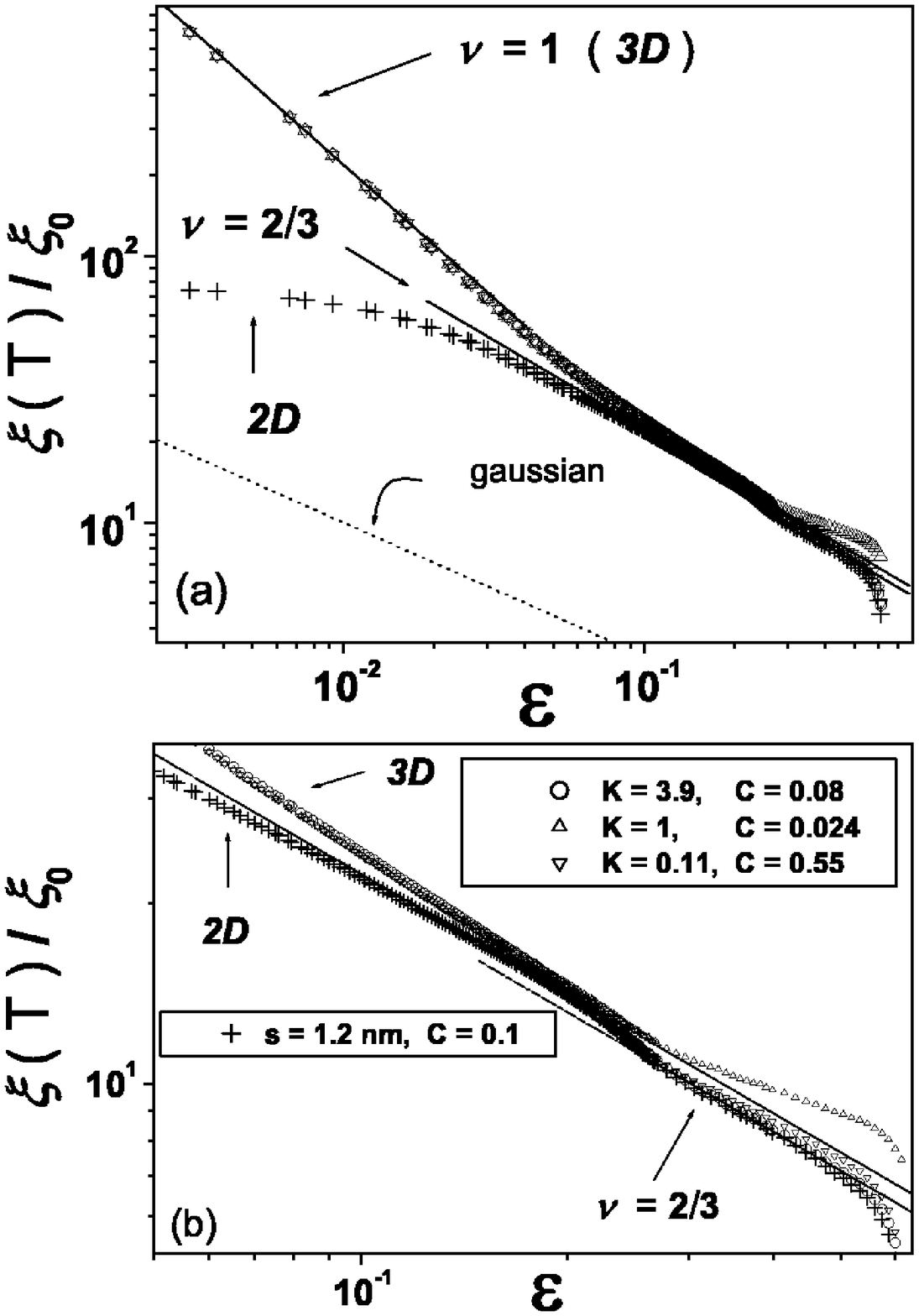}}
\caption{The coherence lengths calculated from $\sigma_2$ in the
various cases indicated in Fig.~\ref{Fig19}. The lower panel shows
the high temperature behavior on an enlarged scale.} \label{Fig20}
\end{figure}

If the unsubtracted data set with $\sigma_1 = \sigma_1^{exp}$ is taken,
one gets the best fit using the anisotropic $\it 3D$ expression with
$K=3.9$ and $C=0.08$ as shown in Fig.~\ref{Fig19}. The other choices for the
cutoff parameters yield $\sigma_1$ curves which depart from the experimental
points much earlier. Also important is to observe that the concomitant
coherence length in Fig.~\ref{Fig20} deviate from the slope 2/3.
\par
As for the $\it 2D$ calculation, one observes that the coherence
length acquires exactly the slope 2/3 at $\epsilon = 0.1$, and the
calculated $\sigma_1^{\it 2D}$ is found on the data points in
Fig.~\ref{Fig19}. The parameter $s = 1.2$nm was used in the $\it
2D$ calculations in this case. The $\it 3D$-$\it 2D$ crossover is
expected at $\xi_c(\epsilon^*) = s/2 = 0.6$nm. From
$\sigma_2(T_c)$ one finds $\xi_{0c}=0.016$ nm in BSCCO-2223 so
that $\xi_c(\epsilon^*)/\xi_{0c} = 38$, and finally from
Fig.~\ref{Fig20} one can evaluate $\epsilon^{*} = 0.05$.
Fig.~\ref{Fig19}(c) shows on an enlarged scale that the calculated
$\sigma_1^{\it 2D}$ fits very well to the experimental points at
that temperature. Hence, the $\it 3D$-$\it 2D$ crossover is a
likely scenario in BSCCO-2223.
\par
An enlarged view of the coherence length at higher temperatures in
Fig.~\ref{Fig20}(b) reveals that there is a step in the temperature
dependence of the coherence length. The origin of this feature is not
yet known. We may only speculate that it is due to the multiple layer
structure of BSCCO-2223.
\par
Also important is to notice in the behaviour of $\sigma_1$ in
Fig.~\ref{Fig19}(a) that the normal conductivity  starts to recover its full
value at higher temperatures than in BSCCO-2212. It means that the pseudogap
is fully opened up to a relatively higher temperature. This will be
discussed in detail in a subsequent section.
\section{YBCO} \label{Sect6}
\begin{figure}
\centerline{\includegraphics[width=0.8\textwidth]{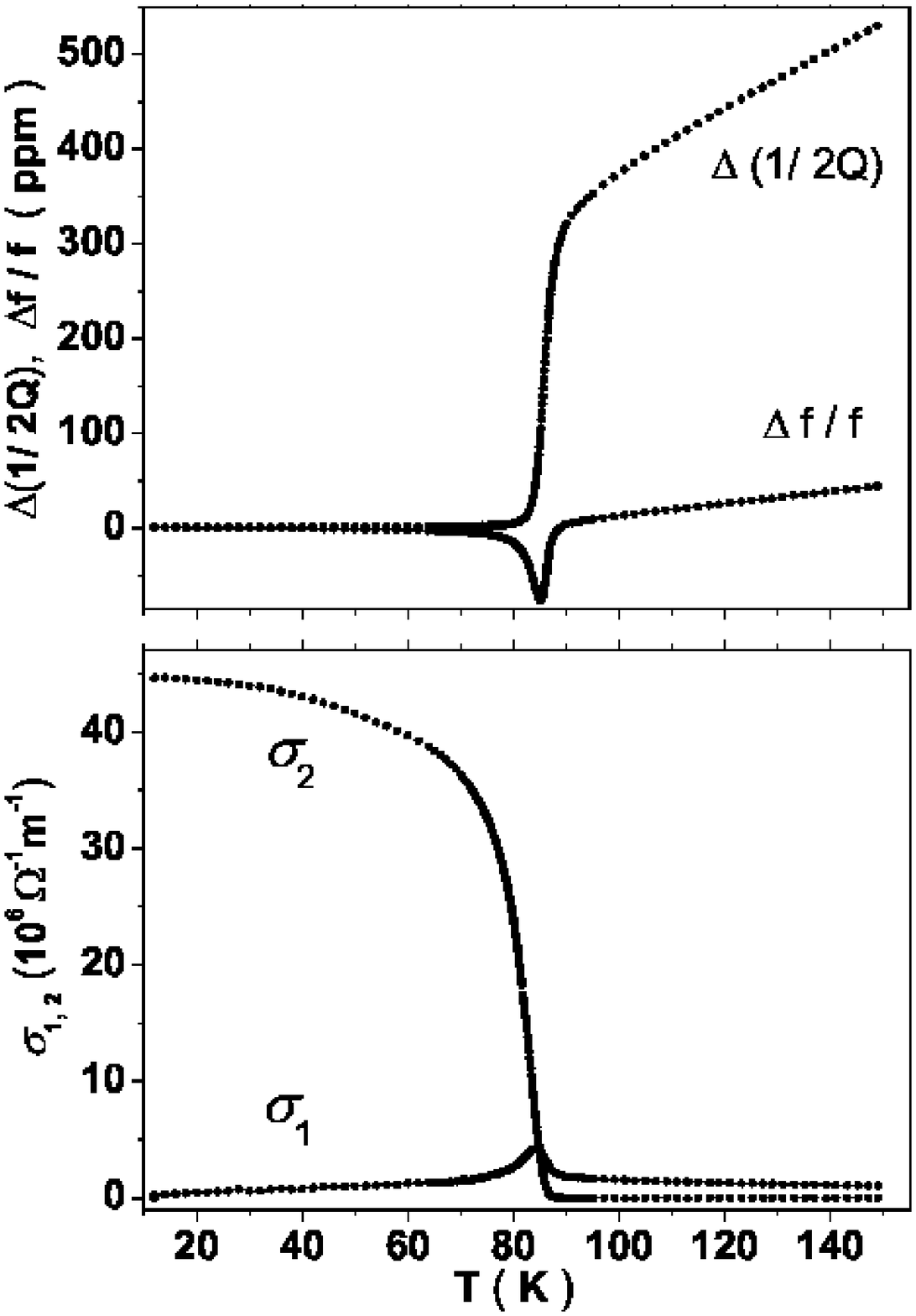}}
\caption{Experimental complex frequency shift and conductivity in
YBCO thin film} \label{Fig21}
\end{figure}

\begin{figure}
\centerline{\includegraphics[width=0.8\textwidth]{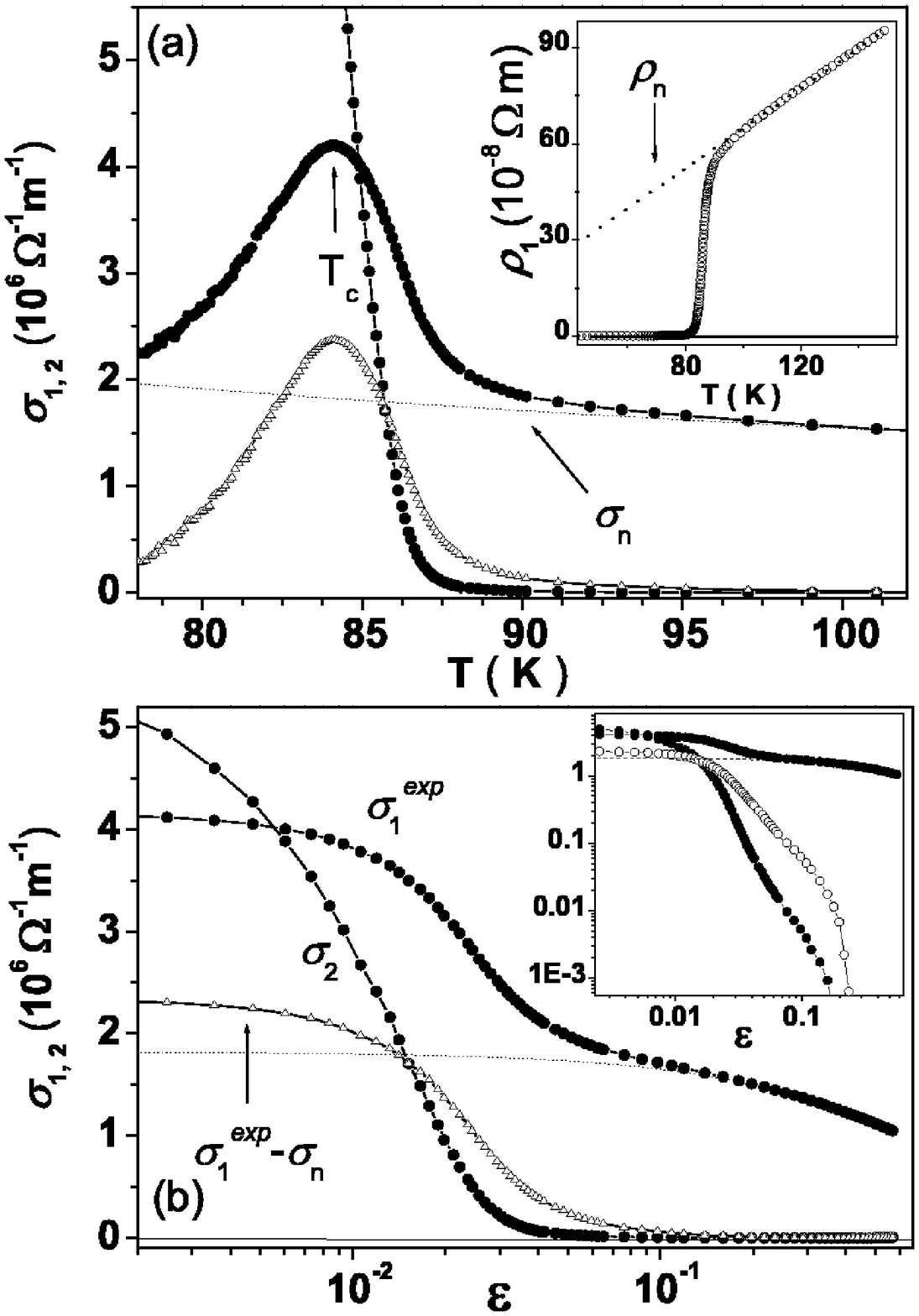}}
\caption{(a) Complex conductivity in YBCO near $T_c$. The
extrapolated linear resistivity $\rho_n$ is shown in the inset.
The as measured $\sigma_1^{exp}$ ($\bullet$) and subtracted
$\sigma_1^{exp}-\sigma_n$ ($\triangle$) data sets are shown. (b)
Fluctuation conductivities above $T_c$ as functions of the reduced
temperature $\epsilon = \ln{\left(T / T_c\right)}$.} \label{Fig22}
\end{figure}

The experimental complex frequency shift and microwave conductivity in
YBCO thin film is shown in Fig.~\ref{Fig21}. The transition appears to be
sharper than in BSCCO thin films. This already points that the fluctuations
in YBCO are weaker. The conductivities are shown in
Fig.~\ref{Fig22} with the two data sets. One can observe that the ratio
$\sigma_2(T_c)/\sigma_1(T_c)$ varies between the two sets much more than
in the case of BSCCO samples. Also one observes that $\sigma_2$ drops much
faster in YBCO. Already at $\epsilon = 0.2$ it has decreased by four orders
of magnitude from its value at $T_c$, and then turns into
noise around zero value. Therefore, the coherence length can only be
calculated up to this temperature. The same holds for the
calculation of $\sigma_1$ as shown in the various cases below.
\par
\begin{figure}
\centerline{\includegraphics[width=0.8\textwidth]{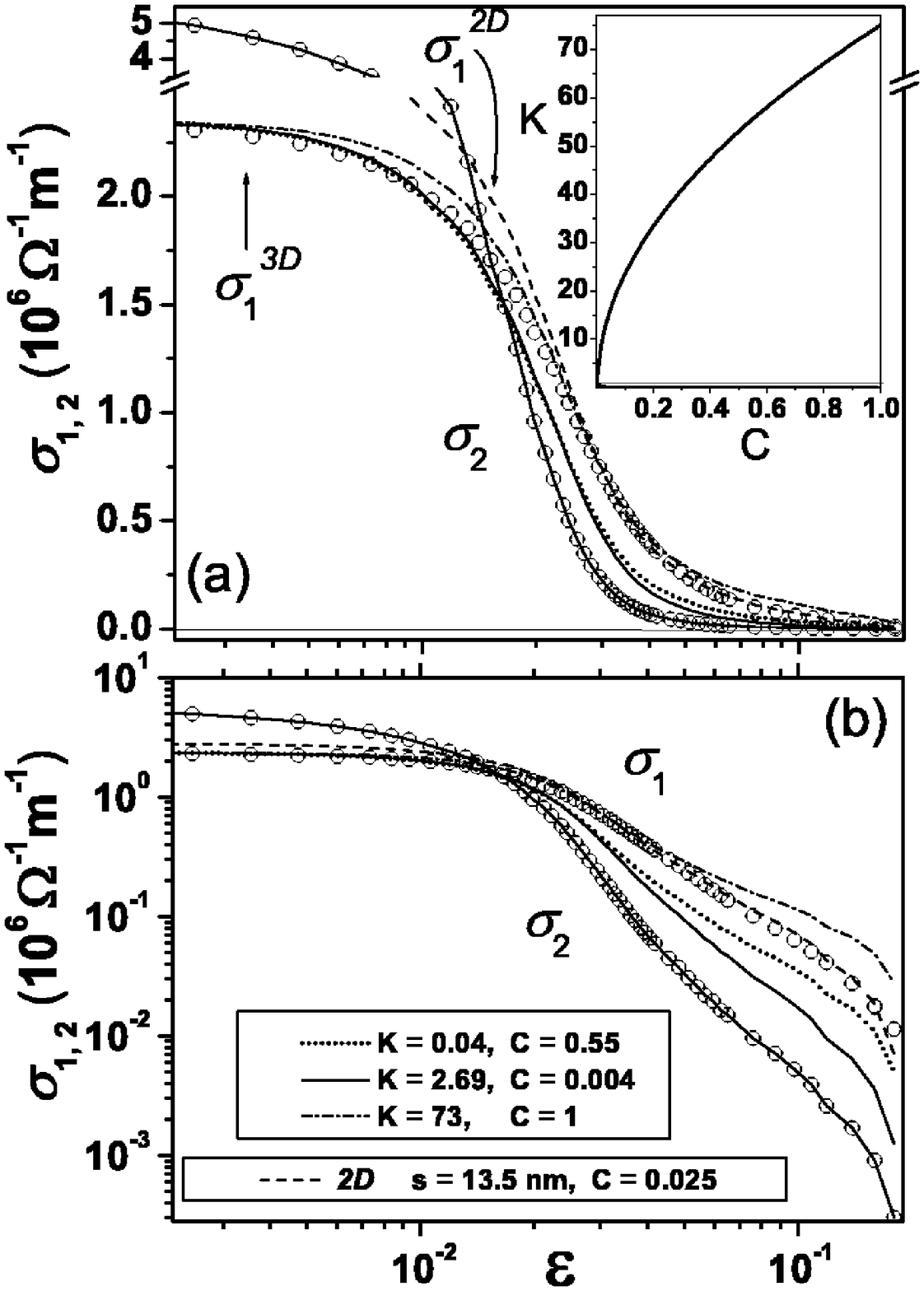}}
\caption{Data analysis on the subtracted data set
$\sigma_1^{exp}-\sigma_n$ and $\sigma_2$ in YBCO thin film. The
presentation is parallel to that of Fig.~\ref{Fig10}. The inset
shows the possible choices of the cutoff parameters for
$\sigma_2/\sigma_1 = 2.42 $ at $T_c$.} \label{Fig23}
\end{figure}
\begin{figure}
\centerline{\includegraphics[width=0.8\textwidth]{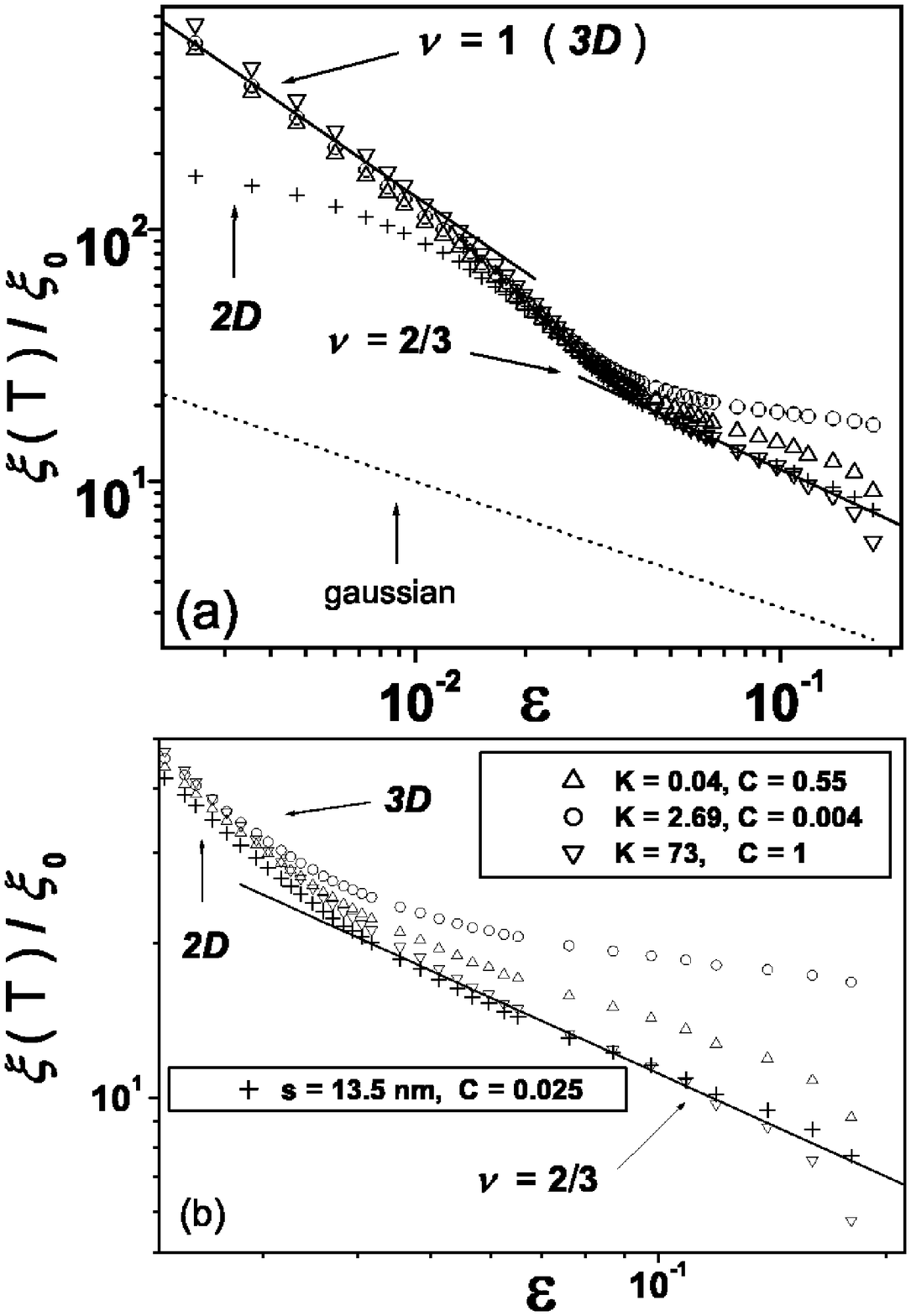}}
\caption{The reduced coherence length calculated from $\sigma_2$
in the cases presented in Fig.~\ref{Fig23}.} \label{Fig24}
\end{figure}

The full analysis based on the data set with subtracted $\sigma_n$ from
$\sigma_1^{exp}$ is presented in Fig.~\ref{Fig23} and the concomitant
coherence lengths are shown in Fig.~\ref{Fig24}. The $\it 3D$ case with
$K \gg 1$ yields $\sigma_1$ values which overestimate the experimental
points at higher temperatures so that one should reject this choice.
The other choices of the cutoff parameters yield lower calculated $\sigma_1$
but the corresponding coherence lengths have slopes which deviate from
the expected 2/3 value. Only the $\it 2D$ coherence length approaches the
slope 2/3, but the fit of $\sigma_1$ in Fig.~\ref{Fig23} is poor so that
even this case has to be rejected.
\par
\begin{figure}
\centerline{\includegraphics[width=0.8\textwidth]{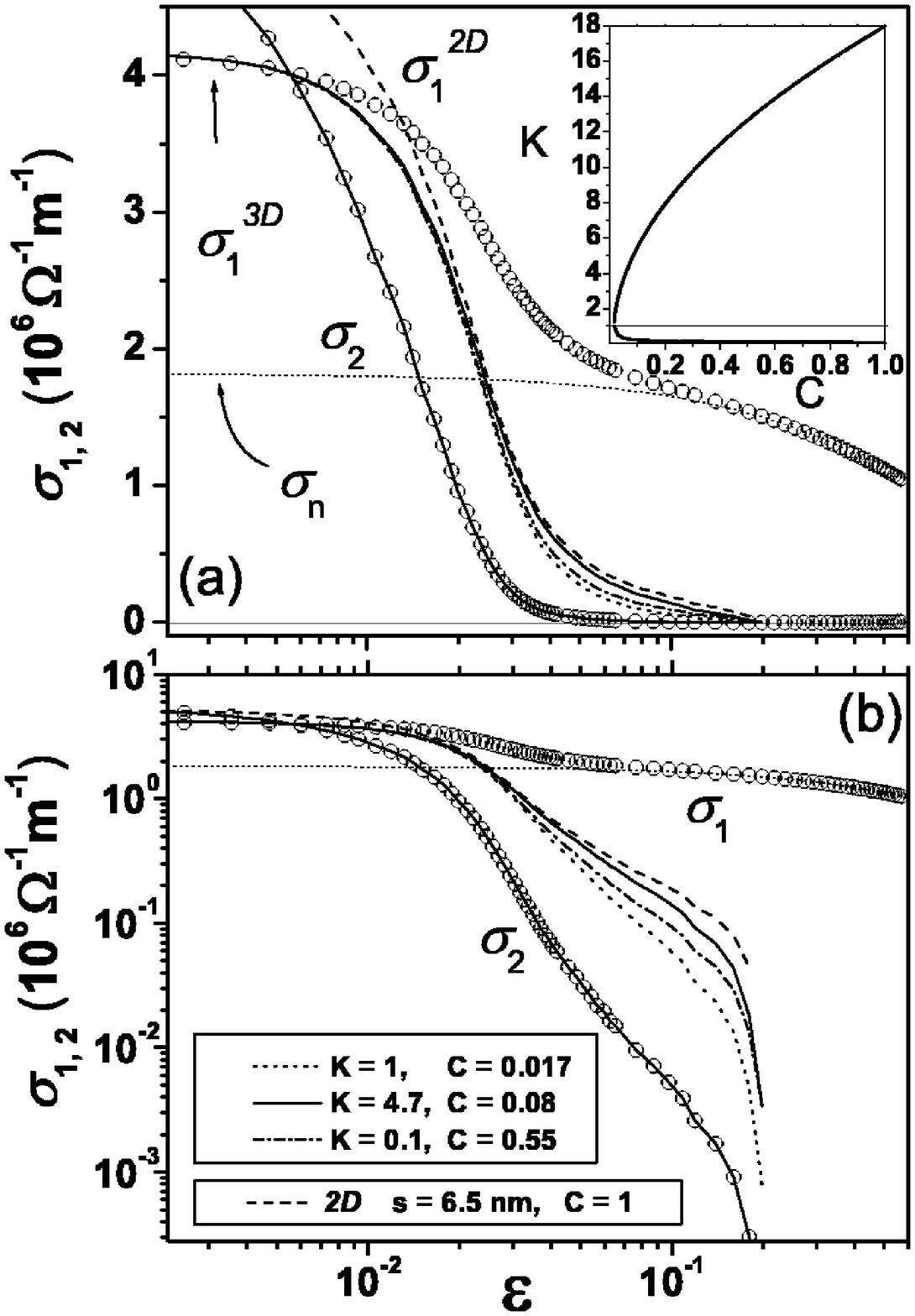}}
\caption{The fluctuation conductivity in YBCO taken as $\sigma_1 =
\sigma_1^{exp}$ and $\sigma_2$ as measured. The inset shows the
possible choices of the cutoff parameters for $\sigma_2/\sigma_1 =
1.36 $ at $T_c$. The calculations were made for the $\it 3D$ and
$\it 2D$ cases as indicated in the legend and described in the
text. The concomitant coherence lengths are presented in
Fig.~\ref{Fig26}.} \label{Fig25}
\end{figure}
\begin{figure}
\centerline{\includegraphics[width=0.8\textwidth]{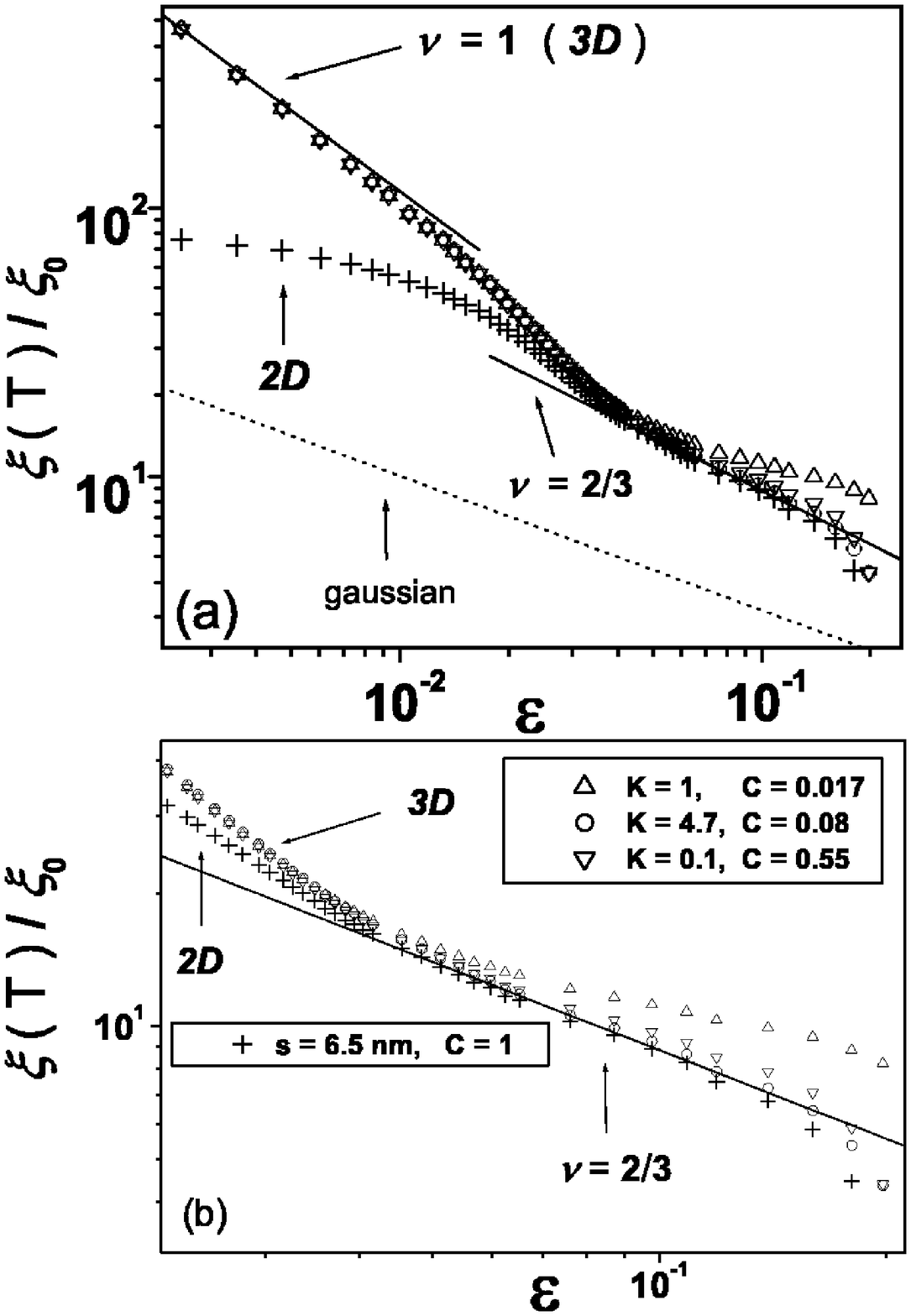}}
\caption{The coherence lengths calculated from $\sigma_2$ in the
various cases indicated in Fig.~\ref{Fig25}. The lower panel shows
the high temperature behavior on an enlarged scale.} \label{Fig26}
\end{figure}

The analysis of the data set where no subtraction was made is
shown in Fig.~\ref{Fig25} and the concomitant coherence lengths
are presented in Fig.~\ref{Fig26}. The best choice of the $\it 3D$
cutoff parameters is $K = 4.7$ and $C = 0.08$. It yields the slope
2/3 for the coherence length at higher temperatures. The
calculated $\sigma_1^{\it 3D}$ deviates early from the
experimental points in Fig.~\ref{Fig25}. This means that the
pseudogap starts to close very soon and the normal conductivity
grows rapidly, reaching its full value already at $\epsilon =
0.2$. Regarding the dimensionality, we find that a $\it 2D$
calculation yields also the slope 2/3 for the coherence length at
temperatures above $\epsilon \approx 0.04$. Since $s=6.5$~nm was
used in this $\it 2D$ calculation, one may expect the $\it
3D$-$\it 2D$ crossover when $\xi_c(\epsilon^*) = s/2 = 3.8$ nm.
From $\sigma_2$ at $T_c$ in YBCO we found $\xi_{0c} = 0.07$ ~nm so
that the crosover should occur at the reduced coherence length
$\xi_c(\epsilon^*)/\xi_{0c} = 47$. Using the data in
Fig.~\ref{Fig26}(a) one finds $\epsilon^* \approx 0.015$. The
calculated $\sigma_1^{\it 2D}$ crosses the experimental points in
Fig.~\ref{Fig25}(a) around this same temperature. Hence, the $\it
3D$-$\it 2D$ crossover in YBCO is not excluded already at
temperatures so close to $T_c$.
\par
The coherence length (Fig.~\ref{Fig26}) shows again the two
critical regimes as in the BSCCO samples, but their temperature
range is squeezed closer to $T_c$. The $3D$ XY critical regime
with $\nu = 2/3$ is found within $0.05 < \epsilon < 0.12$. This
finding is in agreement with the combined analysis of the
$\it{dc}$ conductivity, specific heat, and susceptibility
measurements in YBCO single crystals \cite{Ramallo:99}.
\par
The point of interest is also to look at the absolute values of the reduced
coherence length. Theese are much lower in YBCO than in BSCCO samples at the
same temperature above $T_c$. The observation made above, that the absolute
values of the reduced coherence length are essential for the critical
behaviour, appears to be confirmed. In other words, a given critical regime
persists as long as the reduced coherence length has high enough values.
Due to the divergence of the coherence length in the limit of $T_c$, one
always has to reach this condition. Thus, in classical low temperature
superconductors, the critical regime should also occur, but only within
a tiny temperature interval around $T_c$ which is experimentally unaccessible.
In high-$T_c$ superconductors, however, this temperature interval is very
much extended. As the analysis in the present paper shows, there is a
systematic extension of the critical regime towards higher temperatures
in YBCO, BSCO-2212, and BSCCO-2223 samples.
\section{PSEUDOGAP} \label{Sect7}
\begin{figure}
\centerline{\includegraphics[width=0.8\textwidth]{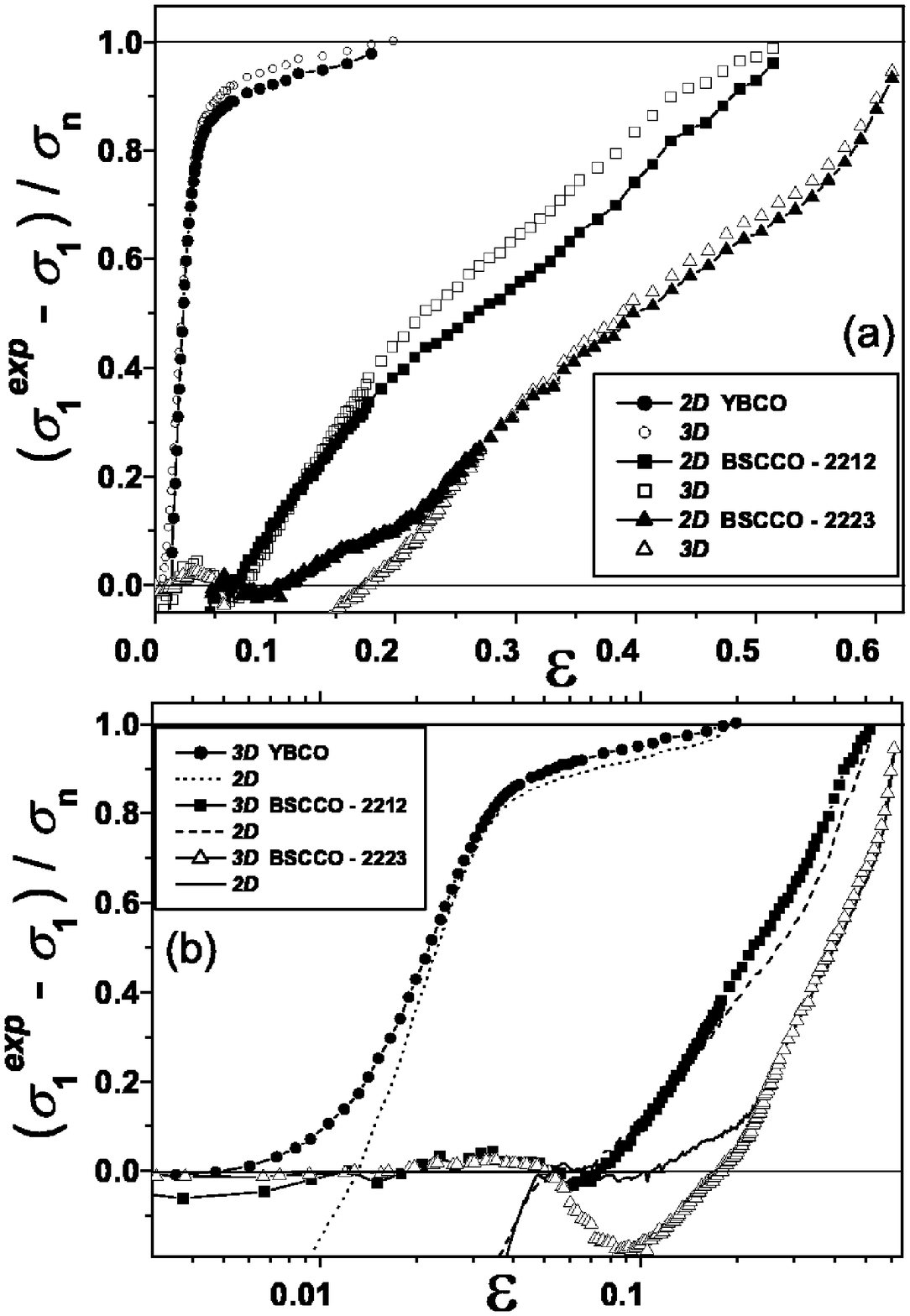}}
\caption{The temperature dependence of the normal conductivity
above $T_c$ in our YBCO and BSCCO thin films. The values are given
as fractions of the normal conductivity extrapolated from the
behaviour at high enough temperatures where the resistivity is
linear.} \label{Fig27}
\end{figure}

It has been shown in the preceding sections that in all our
samples the consistent analysis of the data could be made when no
subtraction of the normal conductivity was made near $T_c$. This
feature is profoundly different from that known in conventional
low temperature superconductors. Obviously, the fact that
superconducting fluctuations are orders of magnitude stronger in
layered high-$T_c$ superconductors, makes their physical behaviour
quite different. It has been observed using other experimental
methods that one electron density of states at the Fermi surface
start to diminish already well above $T_c$. This phenomenon has
been named the pseudogap ~\cite{Timusk:99}. Our analysis of the
microwave fluctuation conductivity has shown that the depletion of
the normal electrons is dramatic near $T_c$. There, the system
obeys the critical regime with $\nu = 1$ and a crossover to the
$3D$ XY critical regime with $\nu = 2/3$. As the latter regime
evolves further by reducing the coherence length, there appears
gradually some normal conductivity which adds to the fluctuation
conductivity to make the experimentally observed one. In
Fig.~\ref{Fig27} we present the growth of this normal conductivity
to its full value $\sigma_n$ in our samples. The presentation is
made with the calculated fluctuation conductivity $\sigma_1$ in
both, $\it 3D$ and $\it 2D$ cases. Only positive values for the
reduced normal conductivity are physically acceptable. Hence, YBCO
is seen to behave certainly as a $\it 3D$ system at temperatures
very close to $T_c$. Ony at $\epsilon > 0.015$ the $\it 2D$
behavior becomes a possible scenario since the calculated normal
conductivity becomes positive. In BSCCO-2212 the $\it 3D$ behavior
is established up to $\epsilon \approx 0.06$. At higher
temperatures the values of the normal conductivity calculated in
the $\it 2D$ case become also positive, and not much different
from the $\it 3D$ case. The interesting situation occurs with
BSCCO-2223. There, the corresponding values calculated in the $\it
3D$ case vanish at temperatures from $T_c$ up to $\epsilon \approx
0.05$, which is the sign of the fully opened pseudogap. In that
temperature region, the $\it 2D$ calculation yields physically
unacceptable negative values. In the interval $0.05 < \epsilon <
0.1$ the $\it 3D$ calculation yields negative values while the
$\it 2D$ calculation now retains the zero values characteristic of
the still fully opened pseudogap. Hence, we conclude that $\it
3D$-$\it 2D$ crossover has taken place at $\epsilon^* \approx
0.05$. At $\epsilon > 0.1$, the $\it 3D$ calculation yields again
positive values for the normal conductivity. At higher
temperatures they become even indistinguishable from those of the
$\it 2D$ case. However, a reversed crossover from $\it 2D$ to $\it
3D$ behavior at higher temperatures is physically unacceptable
because the coherence length does not increase but continues to
shrink to ever smaller values. Therefore we may conclude that the
system remains $\it 2D$ at all temperatures above $\epsilon^*
\approx 0.05$. In view of this observation, we may suggest that
the formal indistinguishability of the $\it 2D$ and $\it 3D$ cases
at higher temperatures should be interpreted in favor of the $\it
2D$ behavior.
\par
It was shown by Corson et al. ~\cite{Corson:99} that underdoped
BSCCO-2212 had extended region of superconducting fluctuations
above $T_c$. The evidence for superconducting fluctuations in
underdoped high-$T_c$ superconductors was found also from the
measurement of the Nernst effect ~\cite{Xu:00,Wang:01}. These
results could support the interpretation of the pseudogap as
preformed Cooper pairs above $T_c$ ~\cite{Emery:95}. However,
direct relationship of the superconducting fluctuations to the
pseudogap could not be made. The relevant study of the pseudogap
features could be achieved by other experimental techniques
~\cite{Timusk:99}. The analysis given in the present paper shows
that both, superconducting fluctuations and the pseudogap, can be
studied by the same experimental technique. A single set of data
can be used to establish a direct relationship of the
superconducting fluctuations to the pseudogap in a given sample.
We find that the two phenomena are indeed intimately related.
Namely, the growing normal conductivity discussed above is the
difference of the experimental values and the calculated
fluctuation conductivity $\sigma_1$. The latter could be
calculated due to the direct measurement of the imaginary part
$\sigma_2$. One should note that the detection of $\sigma_2$ is a
sign of the presence of the superconducting fluctuations. Due to
the high sensitivity of our experimental setup, we could measure
the decay of $\sigma_2$ over four orders of magnitude from $T_c$
up to some sample dependent higher temperature. When this
evolution is compared to that shown in Fig.~\ref{Fig27}, one finds
that the intensity of the superconducting fluctuations is related
to the opening of the pseudogap. In our YBCO sample, the
superconducting fluctuations diminish relatively soon above $T_c$,
and the normal conductivity recovers up to the full $\sigma_n$,
i.~e. the pseudagap closes. In more underdoped BSCCO-2223 sample,
the critical fuctuations extend to much higher temperatures, and
the pseudogap is seen to follow this behaviour concomitantly. We
find this observation to be a strong argument that the loss of the
one electron states at the Fermi level, which constitutes the main
feature of the pseudogap, is due to the strong participation of
the electrons in the superconducting fluctuations.

\section{CONCLUSIONS} \label{Sect8}

We have measured complex microwave conductivity in
$Bi_2Sr_2CaCu_2O_{8+\delta}$, $Bi_2Sr_2Ca_2Cu_3O_{10+\delta}$, and
$YBa_2Cu_3O_{7-\delta}$ thin films above $T_c$. We have found that
the experimental curves for the real and imaginary part of the
$\it ac$ fluctuation conductivity can be consistently interpreted
only if the theoretical expressions take into account a proper
wavevector or energy cutoff in the fluctuation spectrum. Strong
fluctuations extending far above $T_c$ were observed in all of the
three underdoped superconducting films. Our analysis yields the
temperature dependence of the coherence length. Quite
surprisingly, we observe multiple critical regions. Near $T_c$ the
static critical exponent appears to be $\nu = 1$. Following a
crossover, one finds $\nu = 2/3$ at higher temperatures. This
evolution is paralleled by closing of the pseudogap as seen
directly in our analysis through the contribution of the normal
conductivity in the total experimentally observed one. %%

\acknowledgments D.-N.~Peligrad and M.~Mehring acknowledge support
by the Deutsche Forschungsgemeinschaft~(DFG) project
Nr.~Me362/14-2. A.~Dul\v{c}i\'{c} acknowledges support from the
Croatian Ministry of Science. D.~-N.~Peligrad acknowledge also V.
R\v{a}dulescu for the help in the development of some of the
required software for data analysis.

\bibliographystyle{apsrev}

\newpage
\clearpage


\end{document}